%% file: main.tex
\documentclass[sigconf]{acmart}

\AtBeginDocument{%
  }

\copyrightyear{2026}
\acmYear{2026}
\setcopyright{cc}
\setcctype{by}
\acmConference[DIS '26]{Designing Interactive Systems Conference}{June 13--17, 2026}{Singapore, Singapore}
\acmBooktitle{Designing Interactive Systems Conference (DIS '26), June 13--17, 2026, Singapore, Singapore}
\acmDOI{10.1145/3800645.3813005}
\acmISBN{979-8-4007-2563-0/2026/06}
\usepackage{multirow}
\usepackage{stfloats}
\usepackage{array}
\usepackage{framed}
\usepackage{color}

\begin{document}

\title{\sysname{}: Expressing and Reusing Divergent Intents for Graphic Design Exploration using Generative AI}

\author{DaEun Choi}
\orcid{0000-0002-4843-0486}
\affiliation{%
  \institution{KAIST}
  \city{Daejeon}
  \country{Republic of Korea}
}
\email{daeun.choi@kaist.ac.kr}

\author{Kihoon Son}
\orcid{0000-0001-7224-2947}
\affiliation{%
  \institution{KAIST}
  \city{Daejeon}
  \country{Republic of Korea}
}
\email{kihoon.son@kaist.ac.kr}

\author{Jaesang Yu}
\orcid{0009-0001-5779-5305}
\affiliation{%
  \institution{KAIST}
  \city{Daejeon}
  \country{Republic of Korea}
}
\email{jaesangyu@kaist.ac.kr}

\author{HyunJoon Jung}
\orcid{0000-0001-5892-1089}
\affiliation{%
  \institution{MPhora.ai}
  \city{San Francisco, CA}
  \country{United States}
}
\email{hjung@mphora.ai}

\author{Juho Kim}
\orcid{0000-0001-6348-4127}
\affiliation{
  \institution{School of Computing, KAIST}
  \city{Daejeon}
  \country{Republic of Korea}
}
\email{juhokim@kaist.ac.kr}
\affiliation{
  \institution{SkillBench}
  \city{Santa Barbara, CA}
  \country{USA}
}
\email{juho@skillbench.com}

\renewcommand{\shortauthors}{Choi et al.}
\newcommand{\sysname}[0]{IdeaBlocks}
\newcommand{\username}[0]{John}

\newcommand{\obeylineswithdoubleblank}{
  \obeylines
  \let\par\@empty
  \everypar{\ifx\@empty\last@line\par\vspace{0.7em}\fi\gdef\last@line{}}
}

\definecolor{blockcolor}{HTML}{555555}
\definecolor{blockrule}{gray}{0.6}

% \newmdenv[
%   topline=false,
%   bottomline=false,
%   rightline=false,
%   leftline=true,
%   linecolor=blockrule,
%   linewidth=1pt,
%   innertopmargin=2pt,
%   innerbottommargin=2pt,
%   innerleftmargin=10pt,
%   innerrightmargin=0pt,
%   skipabove=10pt,
%   skipbelow=10pt,
%   backgroundcolor=white,
%   font=\small\color{blockcolor}
% ]{block}

% \newmdenv[
%   topline=false,
%   bottomline=false,
%   rightline=false,
%   leftline=true,
%   linecolor=blockrule,
%   linewidth=1pt,
%   innertopmargin=2pt,
%   innerbottommargin=2pt,
%   innerleftmargin=10pt,
%   innerrightmargin=0pt,
%   skipabove=10pt,
%   skipbelow=10pt,
%   backgroundcolor=white,
% ]{blockBig}

% \newmdenv[
%   topline=true,
%   bottomline=true,
%   rightline=true,
%   leftline=true,
%   linecolor=blockrule,
%   linewidth=1pt,
%   innertopmargin=5pt,
%   innerbottommargin=5pt,
%   innerleftmargin=5pt,
%   innerrightmargin=5pt,
%   skipabove=5pt,
%   skipbelow=10pt,
%   backgroundcolor=white,
%   font=\scriptsize\ttfamily\color{blockcolor},
% ]{prompt}

% block
\newenvironment{block}{%
  \def\FrameCommand{{\color{blockrule}\vrule width 1pt}\hspace{10pt}}%
  \MakeFramed{\advance\hsize-11pt\FrameRestore}%
  \small\color{blockcolor}\noindent\ignorespaces%
}{%
  \endMakeFramed%
}

% blockBig
\newenvironment{blockBig}{%
  \def\FrameCommand{{\color{blockrule}\vrule width 1pt}\hspace{10pt}}%
  \MakeFramed{\advance\hsize-11pt\FrameRestore}%
  \normalcolor\noindent\ignorespaces%
}{%
  \endMakeFramed%
}

% prompt (full border, monospace scriptsize)
\newenvironment{prompt}{%
  \def\FrameCommand{\fboxsep=5pt\fcolorbox{blockrule}{white}}%
  \MakeFramed{\advance\hsize-\width\FrameRestore}%
  \scriptsize\ttfamily\color{blockcolor}%
}{%
  \endMakeFramed\vspace{10pt}%
}

\newcolumntype{L}[1]{>{\raggedright\arraybackslash}p{#1}}

\begin{abstract}
\input{Sections/00-Abstract}
\end{abstract}



\begin{teaserfigure}
  \includegraphics[width=\textwidth]{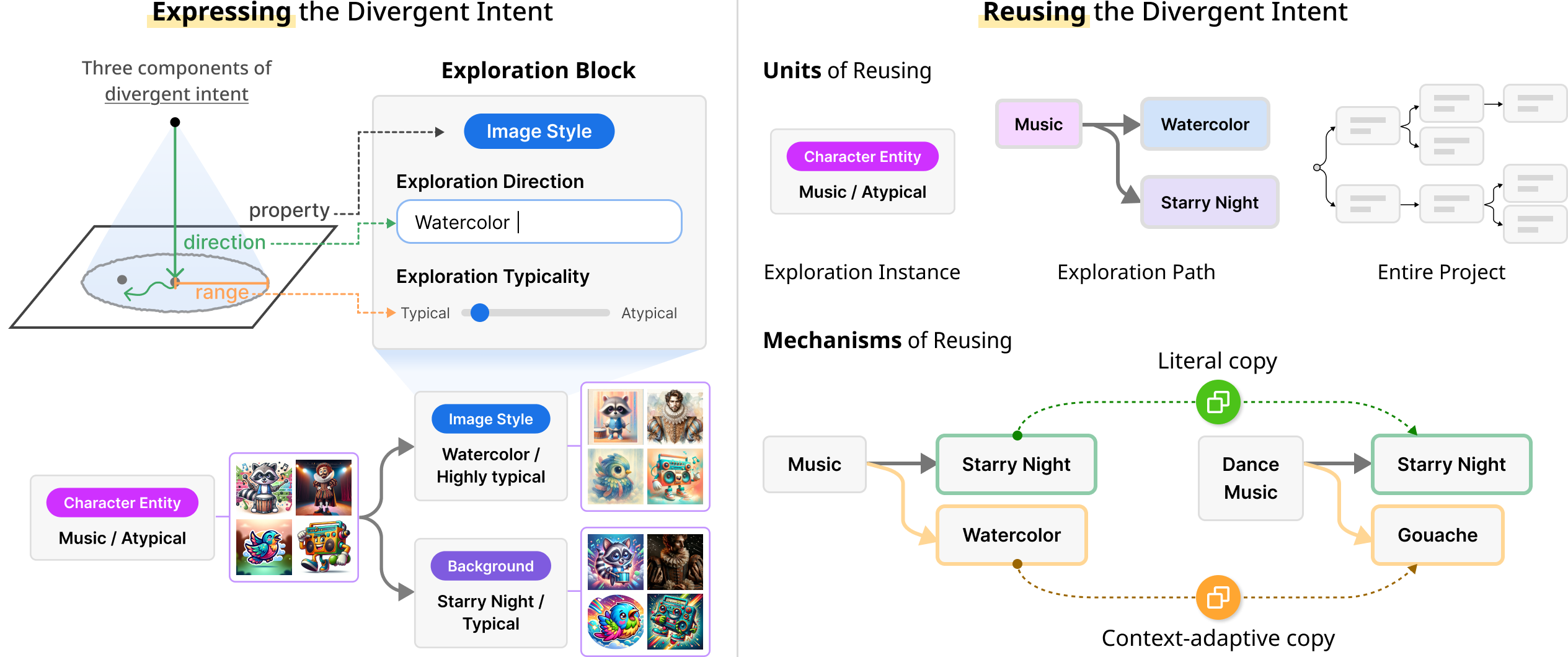}
  \caption{
    Overview of \sysname{}. (Left) Users' divergent intents can be modularized into exploration blocks with three components: property, direction, and range. These blocks can be chained across multiple properties. (Right) Previously created blocks, paths, or entire projects can be reused via literal copy (preserving the original intent) or context-adaptive copy (adapting it to a new context), enabling efficient and iterative exploration.
    }
    \Description{
The figure illustrates the two main components of \sysname{}: expressing and reusing divergent intents.
On the left, the "Expressing" side shows how designers create exploration blocks by specifying three components of intent: the property (e.g., Character Entity, Image Style, Background), the direction (e.g., "Music"), and the range of exploration using a typical–atypical slider. Example outputs demonstrate how varying typicality influences results: a "Music"-related character entity at atypical settings produces unconventional mascots.
Also, the image shows that this block is chained with the following two blocks with a setting of "Watercolor / Highly typical" image style or a "Starry Night / Typical" background, which yields an image that preserves the music-related characters but adds newly explored results. Chaining blocks together allows users to incrementally explore multiple properties in sequence while maintaining prior intent.
On the right, the "Reusing" side illustrates how previously created intents can be reapplied at different scales: a single design property (e.g., Character Entity), an exploration path (a sequence of blocks such as Music → Watercolor → Starry Night), or an entire project (a full exploration graph). Two mechanisms of reuse are supported: literal copy, which directly transfers prior blocks without modification, and context-adaptive copy, which adapts intents to fit a new context (e.g., reusing "Music" with "Starry Night" may adapt into "Dance Music" with "Gouache"). These features allow designers to efficiently branch, remix, and iterate upon past explorations.
    }
  \label{fig:teaser}
\end{teaserfigure}


\maketitle

\input{Sections/01-Introduction}
\input{Sections/02-Related_Works}
\input{Sections/03-Formative_Study}
\input{Sections/04-System}
\input{Sections/05-Evaluation}
\input{Sections/06-Study1}
\input{Sections/07-Study2}
\input{Sections/08-Discussion}
\input{Sections/09-Conclusion}

\begin{acks}
This work was supported by the National Research Foundation of Korea (NRF) grant funded by the Korea government (Ministry of Science and ICT) (No.RS-2025-00557726). This work was also supported by Institute of Information \& Communications Technology Planning \& Evaluation (IITP) grant funded by the Korea government (MSIT) (No.2021-0-01347, Video Interaction Technologies Using Object-Oriented Video Modeling).
\end{acks}

\bibliographystyle{ACM-Reference-Format}
\bibliography{main}

\input{Sections/99-Appendix}

% \clearpage
% \input{Sections/99-Prompts}

\end{document}

%% file: Sections/00-Abstract.tex
While designers increasingly leverage Generative AI for divergent exploration, current interaction is optimized for convergent refinement, forcing users to specify fixed targets rather than open-ended search spaces.
Based on a formative study (N=7), we define the anatomy of \textit{Divergent Intent}, comprising property, direction, and range, and identified two critical barriers: the lack of mechanisms to explicitly shape the parametric boundaries of exploration and the difficulty of reusing successful search strategies.
We present \sysname{}, where users can modularize divergent intents into \textit{Exploration Blocks}. Users can reuse prior intents at multiple levels (block, path, and project) with options for literal or context-adaptive reuse.
In our comparative study (N=12), participants using \sysname{} explored 2.13 times more images with 12.5\% greater visual diversity than the baseline, demonstrating how structured intent expression and reuse support divergent exploration. A three-day longitudinal study (N=6) further revealed how different reuse mechanisms allowed distinct creative strategies, offering design implications for future intent-aware design support tools.

% 164 words

% While designers increasingly leverage Generative AI for divergent exploration, current interaction is optimized for convergent refinement, forcing users to specify fixed targets rather than open-ended search spaces. Based on a formative study (N=7), we define the anatomy of Divergent Intent, comprising property, direction, and range, and identified two critical barriers: the lack of mechanisms to explicitly shape the parametric boundaries of exploration and the difficulty of reusing successful search strategies. We present IdeaBlocks, where users can modularize divergent intents into Exploration Blocks. Users can reuse prior intents at multiple levels (block, path, and project) with options for literal or context-adaptive reuse. In our comparative study (N=12), participants using IdeaBlocks explored 2.13 times more images with 12.5% greater visual diversity than the baseline, demonstrating how structured intent expression and reuse support effective divergence. A three-day deployment study (N=6) further revealed how different reuse mechanisms allowed distinct creative strategies, offering design implications for future intent-aware creativity supports.

%% file: Sections/01-Introduction.tex
\section{Introduction}

Design exploration is essential to the creative process, such as graphic design~\cite{eckert2000sources}, and is characterized by the iteration of divergence and convergence~\cite{finke1996creative, goldschmidt2016linkographic}.
In diverging phases, designers operate with \textbf{\textit{divergent intent}} to experiment widely within a design \textit{space} to \textit{``get the right design''}, while in convergent phases, they adopt \textit{convergent intent} to narrow toward a specific \textit{point} and precisely represent that fixed design idea to validate and \textit{``get the design right''}~\cite{buxton2010sketching}. Divergent intent is not a fixed design goal, but a tentative and revisable orientation for exploration that guides how designers probe and adjust the space of possibilities through interaction with emerging outcomes~\cite{schon1983reflective,dorst2001creativity}.

Generative AI presents a unique opportunity to support divergent intent in graphic design exploration~\cite{ko2023creativeworks}. By enabling rapid generation of visuals from text, it aligns with early-stage prototyping, where low-cost artifacts are used to explore a broad design space~\cite{lim2008anatomy}. However, a structural mismatch remains between this potential and current interaction models.
Existing mechanisms are largely optimized for convergent intent, encouraging users to specify a target outcome through prompt refinement~\cite{brade2023promptify, wang2024promptcharm} or reference images~\cite{shi2025brickify}. This prematurely narrows exploration toward a single point in the design space, which is problematic when designers’ goals are still forming.
Recent systems attempt to broaden exploration by suggesting diverse alternatives~\cite{suh2024luminate, han2025poet}. Yet users still lack mechanisms to \textit{control} their divergence, as they lack parametric means to define the scope and boundaries of their exploration, which are continuous and evolve as designers' intents develop.

To probe this gap, we conducted a formative study with 7 participants, focusing on the challenges of articulating divergent intents when exploring with generative models. From our observation, we identified the anatomy of divergent intent comprising three components---\textit{property}, \textit{direction}, and \textit{range}---which together define the \textit{space} a user intends to explore. However, current generative AI interactions lack mechanisms to express these dimensions, often leading to premature convergence or unintended results.
We further found that once designers discover a satisfactory exploration strategy, it becomes a reusable unit of design knowledge. Rather than reusing individual prompts or outputs, designers seek to reuse the \textit{structure} of exploration—the underlying logic that defines the search space. This suggests that divergent intent reuse operates at a higher level, capturing parametric exploration logic rather than single instantiated prompts.

We propose \sysname{}, a system that enables users to structure their divergent intents during early-stage exploration into modular units called \textit{Exploration Blocks}, based on three components of \textit{property}, \textit{direction}, and \textit{range}.
For instance, a designer creating a mascot character might create a block for \textit{Character Entity}, set the direction to ``Music,'' and adjust the typicality slider toward \textit{Atypical} to explore more divergent variations. Then, they can chain it to other blocks for different design \textit{properties} such as \textit{Image Style} to explore multiple dimensions of their divergent intent simultaneously.
\sysname{} also supports reusing previous divergent intents in three units: block-level (individual exploration instance), path-level (sequence of blocks), and project-level (entire exploration process). At each level, users can choose to reuse intents exactly as before or adaptively in a new exploration context.

We conducted a within-subjects comparative study with 12 designers to evaluate how modularizing and reusing divergent intents in \sysname{} supports divergent design exploration.
The results demonstrate that \sysname{} enhanced participants' ideational fluency and the diversity of their exploration outcomes. They tried 1.77 times more inputs to the model, facilitated by the structured expression of their divergent intent, resulting in 2.13 times more generated images explored within the same time. Their image generation rate increased over time, indicating more sustained exploration driven by active reuse of previous intents. They also explored 12.5\% more visually diverse outcomes, measured by CLIP-based embedding distances across the generated images.
Linkography analysis~\cite{goldschmidt2014linkography} revealed that their explorations were more structured and interconnected with higher entropy, indicating \sysname{} supports broader, more structurally diverse, and less linear idea development.

To further investigate how intent-reuse features of \sysname{} shape design exploration in practice, we conducted a three-day longitudinal study with 6 designers.
For block-level reuse, literal reuse of visual properties such as image styles acted as ``personal palettes'', while semantic properties such as entities or poses were more often adapted to context.
Path-level reuse helped participants recognize their own recurring exploration strategies and reuse them as template-like scaffolds.
At the project level, reuse served two purposes: bootstrapping (to avoid blank starts by building on small, understandable projects) and broadening options (drawing inspiration from others' intermediate strategies).
From these observations, we suggest design implications for future systems to better incorporate intent-reuse features for creative exploration.

Our contributions are threefold:

\begin{itemize}
    \item Formative study findings providing an understanding of the limitations of current generative AI in supporting divergent intent, and a conceptual framework that decomposes divergent intent into three components: \textit{property}, \textit{direction}, \textit{range}.
    \item \sysname{}, which externalizes divergent intents into manipulable and reusable units, demonstrated to be effective through a comparative study.
    \item Design implications for supporting intent reuse in design exploration, derived from a longitudinal study that reveals how and why reuse features are employed across different contexts.
\end{itemize}

%% file: Sections/02-Related_Works.tex
\section{Related Work}

We review related research on (1) generative AI tools that support design exploration and (2) systems that model and reuse user intent in generative workflows.

\subsection{Generative AI Tools for Design Exploration}

\subsubsection{Supporting Divergent Exploration via Generation and Scaffolding}

One strategy to solve ill-defined design problems is to generate diverse variations to expand the solution space~\cite{simon1996sciences}. 
Research has shown that exploring these variations \textit{in parallel}, rather than serially refining a single concept, leads to more divergent outcomes and reduces fixation~\cite{terry2004variation, dow2010parallel}. 
Tools such as Gem-Ni~\cite{zaman2015gem} and Mixplorer~\cite{kim2022mixplorer} automate this by creating parallel variations to broaden potential outcomes.
Similarly, Koyama et al.~\cite{koyama2022bo} asynchronously suggest 
unexplored design candidates using Bayesian optimization.
Another approach facilitates divergence by synthesizing novel outputs through the recombination of distinct sources, as seen in CreativeConnect~\cite{choi2024creativeconnect}, PopBlends~\cite{wang2023popblends}, and Artinter~\cite{chung2023artinter}. Dreamsheets~\cite{almeda2024prompting} utilized spreadsheet-based prompt permutations to generate volumes of variations.

Another approach is to help users widen the design space to overcome fixation~\cite{dow2010parallel}. Systems like Luminate~\cite{suh2024luminate}, MetaMap~\cite{kang2021metamap}, and Dream Lens~\cite{matejka2018dreamlens} help users navigate large generated design spaces — through hierarchical taxonomies, semantic attributes, or interactive visualization of large-scale design datasets. Similarly, POET~\cite{han2025poet} and DesignAID~\cite{cai2023designaid} utilize LLMs to reason about the problem space and suggest diverse design directions or keywords that users might not have considered.

While these systems successfully broaden the options to foster divergent thinking, they rely on selection-based interactions, where users choose from a broadened taxonomy or list of generated candidates.
% This approach focuses on navigating a pre-populated design space, rather than enabling users to explicitly define the exploration they want.
In contrast, \sysname{} complements this approach by giving users direct control over the direction and range of variation, enabling them to actively shape the exploration space itself.

\subsubsection{Managing and Structuring the Exploration Process}

Design ideation is non-linear and often involves branching, merging, and revisiting ideas. Users' exploration strategies vary depending on the maturity of their ideas~\cite{xu2024ideacentric}.
Therefore, tools have been helping users manage and navigate their exploration process. One line of work is to use moodboards~\cite{eckert2000sources, cassidy2011mood}, as used in the systems like Funky Wall~\cite{lucero2015funky}, SemanticCollage~\cite{koch2020semanticcollage}, May AI~\cite{koch2019may}, DesignPrompt~\cite{peng2024designprompt}, and MoodCubes~\cite{ivanov2022moodcubes}. Beyond moodboards, Charrette~\cite{oleary2018charrette} supports designers in curating iteration histories and navigating previous design decisions alongside meeting notes.
More recent attempts are to use a graph-based interface to represent those exploration paths. Expandora~\cite{choi2025expandora} incorporated a mindmap interface to navigate prompt-generated ideas, and Graphologue~\cite{jiang2023graphologue} and Sensecape~\cite{suh2023sensecape} visualize LLM interactions as node-link diagrams. Other systems, including DeckFlow~\cite{croisdale2023deckflow}, XCreation~\cite{yan2023xcreation}, Spellburst~\cite{angert2023spellburst}, and OptiMuse~\cite{zhou2024understanding}, adopt graph-based interfaces that let users express, organize, and reflect on ideation paths through spatial structuring and iteration.

Supporting non-linear workflows aligns closely with the iterative nature of the creative process. Inspired by this, \sysname{} was also designed as a canvas-based node-link system to allow users to interact with their modularized intents in a non-linear manner.

\subsection{Intent Modeling and Reusing in Creative Systems}

As generative AI becomes more integrated into creative workflows, designers often express intentions through text prompts. However, text modality can often be limiting, especially in open-ended ideation where intentions are vague or evolve dynamically. Recent studies investigate these struggles~\cite{mahdavi2024ai, zamfirescu2023johnny}, highlighting the need for systems that better support the expression, resolution, and reuse of creative intent.

\subsubsection{Expressing and Resolving Creative Intents}

Recent research has introduced systems to help users express creative intent more effectively for generative models.
One major approach focuses on prompt refinement, where systems like Promptify~\cite{brade2023promptify}, RePrompt~\cite{wang2023reprompt}, or PromptCharm~\cite{wang2024promptcharm} utilize visualizations or multimodal prompt engineering to iteratively guide users toward a prompt that accurately reflects their mental image. Similarly, PromptMap~\cite{adamkiewicz2025promptmap} and Patchview~\cite{chung2024patchview} provide structured alternatives for prompts to make them more interpretable.
Another approach provides structured input instead of a text prompt. For example, GenQuery~\cite{son2024genquery} and Opal~\cite{liu2022opal} guide structured searches of visual concepts. DesignPrompt~\cite{peng2024designprompt} and StyleFactory~\cite{zhou2024stylefactory} incorporate multimodal inputs to capture users' aesthetic preferences.

However, in design exploration settings, users often begin with open-ended ideas with high uncertainty regarding their goals~\cite{terry2002recognizing, shneiderman2007creativity}, inevitably leading to the use of vague prompts.
To address this, a recent approach incorporates proactive interaction to resolve ambiguity. Hahn et al.~\cite{hahn2024proactive} propose methods to visualize the model's understanding and ask clarification questions to specify user intent. CARE~\cite{peng2025navigating} employs a multi-agent framework for personalized query refinement, and Stylette~\cite{kim2022stylette} interprets high-level styling goals into specific low-level CSS properties.

While prior work effectively helps articulate prompts, its goal is to reduce ambiguity and guide generation toward a specific outcome, which fundamentally supports convergent intent. In contrast, \sysname{} embraces the inherent uncertainty of early-stage ideation not as a problem to be solved, but as a parameter to be controlled. Rather than attempting to fix the vague intent, \sysname{} allowed users to explicitly express divergent intent by defining the direction and range of variance.

\subsubsection{Reifying and Reusing Creative Intents}

Another line of research explores how users' intents can be retained and reused. Drawing from the instrumental interaction model~\cite{beaudouin2000instrumental}, recent frameworks have proposed reifying interactions with AI into manipulable objects~\cite{kim2023cells, mackay2025interaction}.
AI-Instruments~\cite{riche2025ai} and Memolet~\cite{yen2024memolet} turned user prompts or chat interactions into reusable widgets or memory objects to enhance operational efficiency. IntentTagger~\cite{gmeiner2025intent} introduces micro-prompts for granular specification that can be reused during the co-creation process. Not only for the piece of text prompt, but the recent systems, such as Brickify~\cite{shi2025brickify}, create persistent design tokens from images to facilitate the reuse of specific visual attributes.

Prior approaches largely focus on reusing static content, operational commands, or local specifications. Motif~\cite{kim2015motif} reuses expert creative patterns to scaffold novice work, and SelPh~\cite{koyama2016selph} progressively learns users' personal preferences from editing history to guide future actions. More recently, Brickify~\cite{shi2025brickify} reuses specific visual properties to ensure consistency, while AI-Instruments~\cite{riche2025ai} reuses functional controls for efficiency, and both primarily support convergent objectives. In contrast, \sysname{} reifies the exploration logic by modularizing the strategy for probing the design space, allowing designers to transfer their divergent intent across different contexts rather than simply repeating fixed actions.

\begin{figure*}
    \centering
    \includegraphics[width=\linewidth]{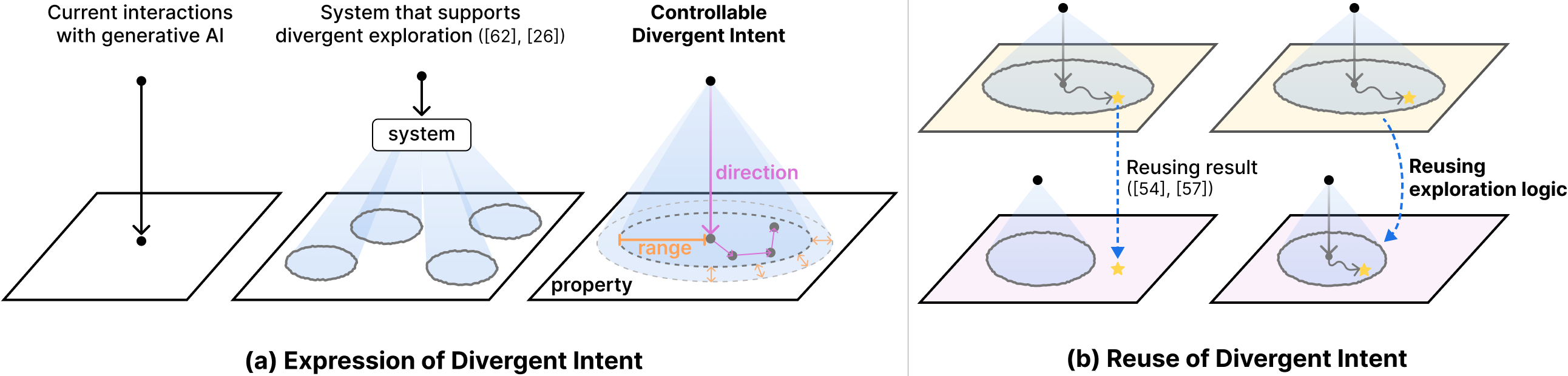}
    \caption{
    Conceptual distinction between \sysname{} and prior works. (a) Expression: Current GenAI interactions are optimized for specifying a single point. Prior research adds divergence but lacks user control over its boundary. \sysname{} enables users to shape parametric boundaries (Property, Direction, Range), freely navigating a property space and controlling the range of variation. (b) Reuse: Prior methods reuse static assets or prompts. \sysname{} reuses the exploration vector (Direction + Range), decoupling the strategy from the content to transfer successful exploration logic across design contexts.}
    \Description{
    This figure shows a two-part conceptual diagram distinguishing IdeaBlocks from prior work. (a) Expression compares three interaction models in a 2D design space: 'Current GenAI' shows a narrow path converging on a single point; 'Prior Research' depicts discrete points scattered broadly but randomly; 'IdeaBlocks' illustrates a controlled, fan-shaped search space bounded by explicit parameters labeled Property, Direction, and Range. (b) Reuse compares two transfer mechanisms between 'Context A' and 'Context B'. 'Prior Methods' shows Coordinate Reuse, where a static destination point is copied, pulling the result back towards the original context. 'IdeaBlocks' shows Vector Reuse, where an arrow representing the exploration strategy (Direction and Range) is lifted from Context A and applied to Context B, guiding the search to a new, context-appropriate area.
    }
    \label{fig:divergent_intent}
\end{figure*}

%% file: Sections/03-Formative_Study.tex
\section{Formative Study}

To better understand the limitations of using current generative AI tools for early-stage design exploration, we conducted a formative study, especially focused on the challenges when expressing and evolving their divergent intent.
We recruited 7 participants (6 females; age M = 24.86 and SD = 3.34) through an online recruitment post. All participants had prior experience with image generation models for design tasks and had worked on at least three graphic design projects before. Details about the participants are in Appendix~\ref{appendix:formativeparticipant}. The study lasted for an hour, and participants were compensated 25,000 KRW (approx \$17 USD).

\subsection{Study Procedure}
Participants were asked to explore diverse mascot character designs for their company (for practitioners) or department (for students) with image generation models for 40 minutes. This task was chosen for its open-ended nature, which encouraged divergent thinking, while also being sufficiently familiar to all participants with different design experiences. Following the session, we conducted a 15-minute semi-structured interview.

We provided a custom image-generation interface to facilitate think-alouds about their intent during exploration.
The system supported four methods for image generation: (1) text-to-image prompting, (2) image-to-image generation with text guidance, (3) text-guided inpainting, and (4) generating variations from an image.
We used Stable Diffusion XL~\cite{podell2023sdxl} for text-to-image, InstructPix2Pix~\cite{brooks2023instructpix2pix} for image-to-image, and DALL·E 2 API~\footnote{https://openai.com/index/dall-e-2/} for inpainting and variations.
Whenever participants use one of them, the interface displays a pop-up message asking them to articulate their current intents and expectations. After reviewing the generated results, the tool prompted them to reflect on whether the results accurately reflected their intention, and if not, what they thought the reason was. 
In total, we collected 158 data points, each consisting of a participant's expressed intent, system action, and participant's assessment of how well the outputs aligned with their intent.

We conducted an inductive thematic analysis of the collected data and interview data. Open coding was performed iteratively after each session until thematic saturation was reached.

\subsection{Findings}
\subsubsection{Characterizing Divergent Intents in Design Exploration}
Design exploration begins with an open-ended intent to explore the \textit{space} of possibilities, rather than a single \textit{point}.
We observed that participants rarely articulated such intent as a fixed design goal, but in terms of how they wanted to orient exploration by specifying where to look, what to vary, and how broadly to explore, while remaining open to revision. This suggests that intent in early-stage exploration functions as a tentative, revisable probe into the design space~\cite{schon1983reflective,dorst2001creativity}, which we define here as \textit{divergent intent}.

Through our thematic analysis of participants' articulation attempts of this exploration space, we identified that \textit{divergent intent} could be structured with three components (Figure~\ref{fig:divergent_intent}). While participants did not use these terms explicitly, their articulations consistently reflected attempts to specify what aspect to vary, in which direction, and how broadly — which we abstract into \textit{property}, \textit{direction}, and \textit{range}, respectively (see Appendix~\ref{appendix:codebook} for the full codebook, including representative quotes and examples from participants).

\begin{itemize}
    \item \textit{Property}: Dimension; \textit{What} aspect to explore (e.g., entity of a mascot, background setting, image style)
    \item \textit{Direction}: Center; \textit{Where} to center the exploration within the property (e.g., ``a cat mascot'' for entity, ``watercolor style'' for image style)
    \item \textit{Range}: Magnitude; \textit{How far} to deviate from that center (the breadth of variation of the results)
\end{itemize}

\subsubsection{Mismatch Between Divergent Intent and Convergent Controls}

\paragraph{Lack of Control over Exploration Range.}
As participants initiated their explorations, they frequently shifted between seeking results closely aligned with their original input (convergent) and pursuing options that ventured into atypical territory (divergent). Most prominent challenge was their inability to control this \textit{range} of exploration. Current interfaces do lack explicit controls for the desired breadth of divergence, and to work around this, participants tried to push their prompts with vague or extreme modifiers like ``various styles'' or ``doing something'' in their prompts (P2, P3) or by relying on the variation feature (P1, P3, P5-7). However, these strategies were unreliable, often overshooting into randomness or undershooting into homogeneity, because the system could not interpret the intended magnitude of variance. While recent approaches~\cite{suh2024luminate, han2025poet} have attempted to broaden exploration support by structurally suggesting various design dimensions, they primarily support \textit{categorical} breadth rather than \textit{parametric} variance, preventing users from manipulating the magnitude of divergence along those dimensions (Figure ~\ref{fig:divergent_intent}). Therefore, we need a way to operationalize divergence not as a discrete state, but as a continuous and tunable parameter, allowing users to explicitly calibrate the scope of exploration to match their evolving intent.

\paragraph{Friction in Evolutionary Steering.}

As exploration unfolds, users’ design intentions on what to explore next naturally evolve in response to the results they see~\cite{dorst2001creativity}. Reflecting this, all participants repeatedly refined their prompts after each iteration, yet this process introduced substantial friction in two key ways.
First, users had to mentally translate the visual preferences they observed back into explicit text descriptions. Second, as the text prompt entangles all aspects of the image, modifying the prompt to refine the \textit{direction} of a specific \textit{property} often unintentionally affects other \textit{properties}. Across sessions, participants expressed a need for \textbf{bounded iteration}.

Several participants (P1, P2, P5, P7) envisioned that when starting from a vague direction, the system could suggest candidate directions within the property and range that they're interested in, allowing them to refine the exploration by accepting or rejecting these options. This highlights the need for a lightweight mechanism to ``lock'' the exploration scope and iteratively steer the direction within it.

\subsubsection{Lack of Support for Reusing Past Divergent Intents}
Once participants found an exploration trajectory that suited their evolving intents, they looked for ways to reuse that same exploration pattern to new contexts, applying the underlying logic of how they probed the design space, rather than merely reusing the results they liked.

Some participants tried to do this by copying and pasting descriptions from the previous prompts (P1, P2, P3). However, participants found that simply reusing parts of a text prompt failed to transfer this logic, as text prompts entangle the content (what to draw) with the exploration parameters (how to vary it). For instance, P1 initially explored ``drawing of a soft cloud'' and appreciated that the generated results helped her see a broad range of diverse drawing styles. Hoping to apply this same diversity of style to a specific character, she reused the phrase ``soft drawing style'' in a new prompt: ``young boy with his head and legs sticking out of a cloud, soft drawing style.'' However, instead of producing the diverse range she liked, the model converged on a homogenous set of typical pencil drawings. Because the user could not explicitly decouple and reuse the strategy of divergence from the text prompt, they lost the exploration space they had found and wanted to reuse.

% TODO: 아래 문단 필요한지 생각해보기
We also observed that the scope of reusing previous intent varied across situations. For example, when participants wanted to reuse divergent intents tied to their global preferences, they wanted to reuse them across different projects. When the intents were tied to more specific ideas, they wanted reuse to remain confined to that idea and to be able to escape it when exploring other ideas.
We conceptualize this variation along two dimensions: \textit{unit}, or how much of the previous exploration is reused, and \textit{adaptivity}, or how fixed versus flexible that reuse should be across different exploration contexts. These two axes form the basis of how \sysname{} supports the reuse of divergent intent.

\subsection{Design Goals}

Based on the findings, we derived two design goals (DGs) for the system to address the challenges faced by designers during the use of the generative models in early-stage design exploration:

\textbf{DG1. Support expressive articulation of divergent intents.}
Designers struggled both to articulate the direction and the range of the divergent intent. The system should provide mechanisms for expressing direction of divergent intents more flexibly, and allow explicit control over how broadly to explore around a given direction.

\textbf{DG2. Enable flexible reuse of past divergent intents.}
Designers often wanted to reuse the aspects of their explorations. Their reuse needs varied across scopes. A system should externalize intents into reusable units that can be retrieved and reused in diverse ways.

%% file: Sections/04-System.tex
\section{\sysname{}}

\begin{figure*}[!t]
    \centering
    \includegraphics[width=1\linewidth]{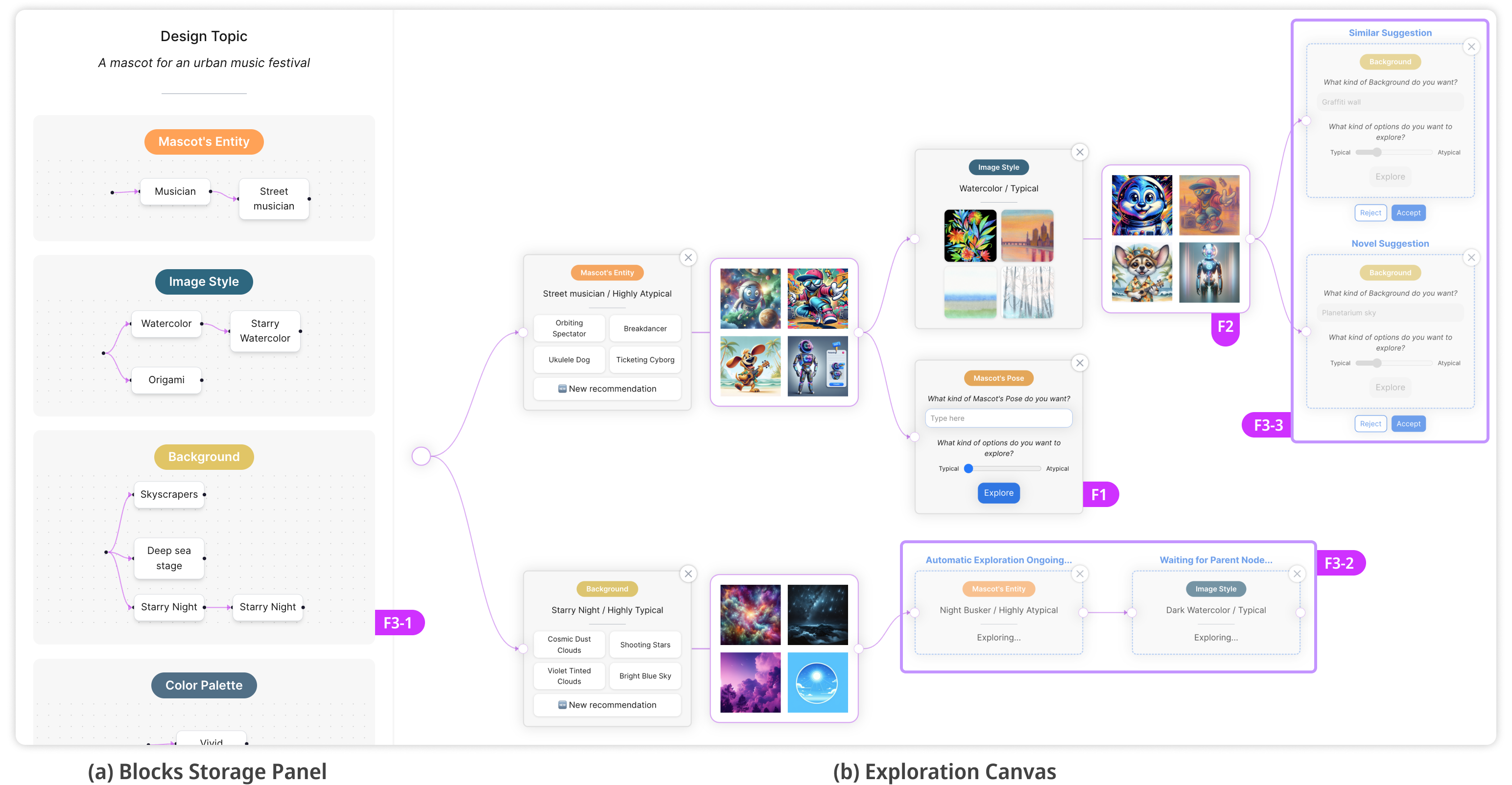}
    \caption{Screenshot of the \sysname{} system with (a) the \textit{Blocks Storage Panel} and (b) the \textit{Exploration Canvas}. Users express their divergent intent through structured inputs in \textit{Exploration Blocks} (F1), which can be chained to build on prior ideas across different properties. Users can reuse individual blocks (F2-1), entire exploration paths (F2-2), or the whole project, with optional adaptation.
    }
    \Description{This figure presents an overview of the \sysname{} system interface, designed to support structured and flexible design exploration workflows. (a) The Blocks Storage Panel on the left shows previously created \textit{Exploration Blocks}, organized by design property such as ``Mascot's Entity'', ``Image Style'', and ``Background''. Users can revisit and reuse these blocks to support efficiency and continuity. (b) The Exploration Canvas on the right is where users conduct active exploration. In F1, users express their divergent intent by creating an \textit{Exploration Block}, specifying a keyword and adjusting the typicality slider. This structured input defines the direction and breadth of the generated suggestions. In F2, users chain together multiple \textit{Exploration Blocks} across properties. Each new block builds upon the content and results of the previous one, allowing users to maintain continuity while refining or expanding their ideas. F3-1 shows how users can reuse a single block from the sidebar and reconnect it to new paths. F3-2 demonstrates the reuse of an entire exploration path, with the option to apply it literally or as a context-adaptive copy that modifies its content to fit the new location. Together, these features enable structured, iterative, and non-linear design exploration across multiple creative dimensions. F3-3 illustrates the system's automatic suggestions for the next exploration direction, offering both familiar and novel options based on prior user activity and the current canvas context.
    }
    \label{fig:system-screenshot}
\end{figure*}

Based on the design goals, we implemented \sysname{} to support designers' early-stage exploration by enabling the expression and reuse of divergent intents. The core of \sysname{} is the concept of an \textit{\textbf{Exploration Block}}---a modular unit that captures the user's divergent intent. Each block (Figure~\ref{fig:feature1}) includes three components identified in our formative study: property (defined by selecting a type of exploration block), direction (specified via a keyword input), and range (controlled through the typicality slider).

To ease the challenge of steering directions, \sysname{} augments user input with system-generated textual and visual suggestions, enabling users to refine directions by selecting, modifying, or extending examples.
To control the range, we intentionally chose \textit{typicality} as a parameter because it captures an implicit but common concern in early-stage exploration --- how far to deviate from a familiar center --- and is easy for designers to reason about across different design properties. Typicality serves as one concrete instantiation of how the range of divergent intent can be externalized and controlled. We discuss how this approach can be expanded in Section~\ref{sec:generalizability}.

Through this structure, \sysname{} enables users to explicitly define their divergent intent and modularize it into \textit{Exploration Blocks} (DG1). These blocks can be chained to preserve prior exploration and build new explorations incrementally. Previously explored blocks and paths are stored as a reusable format, allowing users to efficiently revisit and reuse their past explorations (DG2).

\subsection{Interface Walkthrough \& Features}
\sysname{} provides a canvas interface for creating and interacting with \textit{Exploration Blocks} (Figure~\ref{fig:system-screenshot}). In this section, we illustrate its key features through a scenario with \username{}, a concept artist designing a mascot character for a music festival.

\subsubsection{Generate a List of Properties}

The exploration begins by entering a design topic, and \sysname{} uses GPT-4o to suggest relevant design properties, which are shown in the left panel (Figure~\ref{fig:system-screenshot} (a)). Two types of properties are generated, either text-based (e.g., ``Character's Entity'') or image-based (e.g., ``Image Style''). Users can also add their own properties to explore.

\begin{block}
\username{} wants to design a mascot for the music festival organized by his team, and he inputs the topic into \sysname{}. The system presents a list of properties to explore, including ``Mascot's Entity'', ``Image Style'', and ``Background''.
\end{block}

\subsubsection{Expressing Divergent Intent into Exploration Block}

\begin{figure*}
    \centering
    \includegraphics[width=1\linewidth]{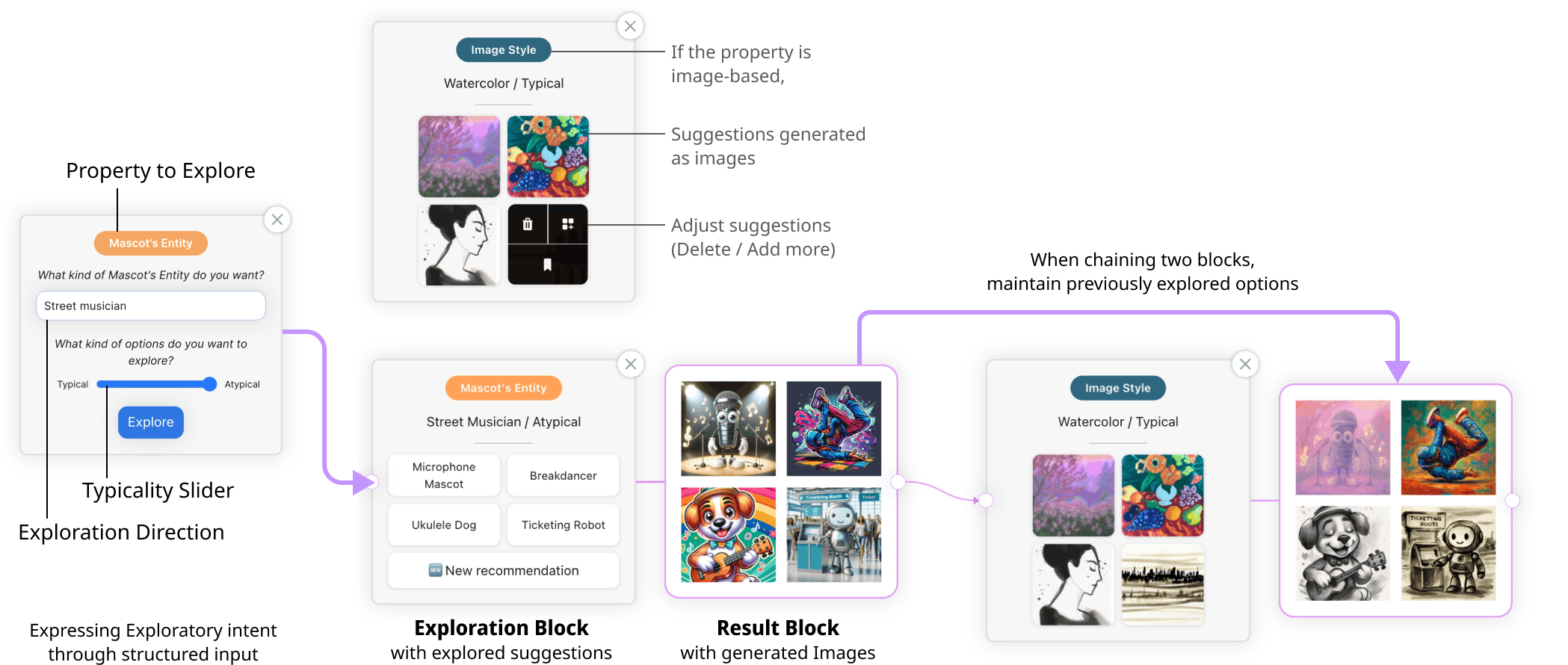}
    \caption{An \textit{Exploration Block} comprises a property type, exploration direction, and typicality slider. Based on these settings, \sysname{} generates four suggestions as text or images, depending on the property (e.g., text for Mascot's Entity, images for Image Style), which users can delete or request more of. The adjacent \textit{Result Block} displays generated images. When chaining blocks, each \textit{Result Block} carries forward previously explored options, allowing users to build on prior intent across properties.}
    \Description{
    Interface for expressing divergent intent in \sysname{}. In this example, the user selects the property ``Mascot's Entity`` and inputs `Street musician` as the exploration direction. The typicality slider is set to a high level (`Highly Atypical`). Based on this input, the system generates four creative suggestions: `Microphone Mascot,' `Breakdancer,' `Ukulele Dog,' and `Ticketing Robot.' These suggestions appear in the \textit{Exploration Block}, where users can delete, bookmark, or request additional ideas. For image-based properties such as ``Image Style'', the system generates visual suggestions directly—for example, four stylistically different watercolor illustrations are shown in the suggestion panel. Along with the suggestions, the system displays the generated images in the adjacent \textit{Result Block}.
    Chaining \textit{Exploration Blocks} to maintain context and continuity in \sysname{}. In this example, the user already has an \textit{Exploration Block} for the ``Mascot's Entity'' property with the direction `Street musician' and a high typicality level. Then, the user creates a new \textit{Exploration Block} for the ``Image Style'' property and chains it to the previous block. The style direction is set to `Watercolor' with a low typicality level. Because the new block is linked to the prior one, the system incorporates both the newly specified style and the previously explored entity options when generating the next set of images. As a result, the user receives a coherent visual outcome that reflects both the diverse `Street musician' character concepts and the typical `Watercolor' styles, enabling a layered and continuous exploration experience.
    }
    \label{fig:feature1}
\end{figure*}

For each design property, users create an \textit{Exploration Block} by entering a keyword and adjusting a typicality slider. The keyword sets the initial direction of exploration (e.g., `Street musician'), and the typicality level (1-5) controls the exploration breadth: lower typicality levels generate more aligned and conventional results, while higher levels produce more divergent outputs.

Based on these inputs, the system generates four suggestions, shown as text or images depending on the property type. Technical details for generating suggestions are presented in Section~\ref{sec:tech1}. Users can delete undesired suggestions or request similar or more distant suggestions, and the system updates exploration direction based on each user action.
The images generated from each suggestion appear in a \textit{Result Block} next to its corresponding \textit{Exploration Block}.

Each \textit{Exploration Block}, along with its parameters and outcomes, is stored and summarized in the left panel. These stored blocks can later be reused for further exploration, as described in Section~\ref{sec:reuse}.

\begin{block}
\username{} begins by exploring ``Mascot's Entity'' and creates an \textit{Exploration Block} for this property. He enters `Street Musician' as the initial direction and sets the typicality level to 5 to explore broadly and discover unexpected ideas. The system generates suggestions such as `Ukulele Dog' and `Ticketing Robot' (left side of Figure~\ref{fig:feature1}). \username{} refines the direction by removing irrelevant ones and getting more suggestions around preferred ones. He then reviews the images generated in the \textit{Result Block}.
\end{block}

In addition to manually entering a keyword, users can receive system-generated direction recommendations, which draw on the user's prior exploration history and the current context (i.e., connected blocks and property type). Using GPT-4o, the system provides both familiar and novel directions.

\subsubsection{Chaining Multiple Blocks}

Users can chain multiple \textit{Exploration Blocks} to build on earlier ideas across different properties. When a new block is linked to an existing one, its setting, suggestions, and generated results become input for the next stage of exploration, and images generated in subsequent \textit{Result Blocks} reflect both the current intent and the accumulated exploration history. Technical details of this image generation process are in Appendix~\ref{appendix:tech}.

\begin{block}
After exploring different character entities, \username{} moves on to the next property, ``Image Style''. He creates a new \textit{Exploration Block} and connects it to the previous one. He explores typical watercolor styles and adjusts the system's suggestions, just as he did with the ``Mascot's Entity''. The generated images shown in the \textit{Result Block} reflect not only the watercolor style, but also incorporate the previously explored character entities from the connected block (right side of Figure~\ref{fig:feature1}).
\end{block}

\subsubsection{Reusing Previous Intentions} \label{sec:reuse}

\begin{figure*}
    \centering
    \includegraphics[width=1\linewidth]{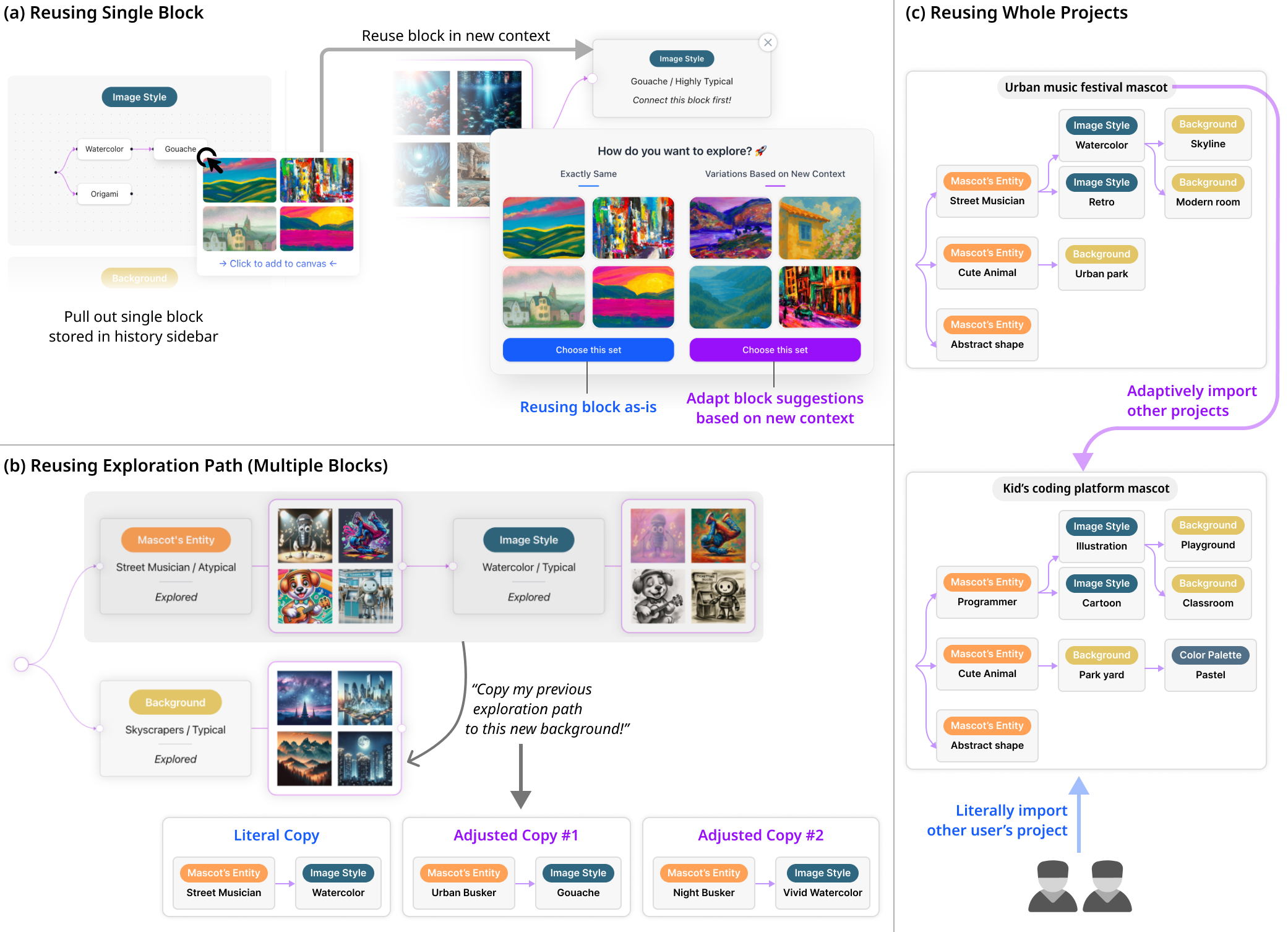}
    \caption{\sysname{}'s four features for reusing prior exploration. (a) A previously created \textit{Exploration Block} can be pulled from the history sidebar and placed on the canvas as-is (literal) or with suggestions adapted to the new context. (b) An entire \textit{Exploration Path} can be copied as a literal duplicate or as context-adaptive variants with adjusted block parameters. (c) A previous project can be adaptively imported to bootstrap a new topic, or another user's project can be literally imported to borrow their exploration structure.}
    \Description{The figure illustrates the three levels of reuse supported in \sysname{}: reusing single blocks, reusing entire exploration paths, and reusing whole projects.
    (a) Reusing Single Block. Users can pull out an individual block (e.g., "Gouache" image style) from the history sidebar and apply it in a new context. The system offers two options: reuse the block exactly as-is (literal copy) or adapt its suggestions based on the new context (context-adaptive copy).
    (b) Reusing Exploration Path. Users can copy an entire sequence of blocks (e.g., "Street Musician / Atypical" → "Watercolor / Typical" → "Skyscrapers / Typical") and apply it to a new background or property. The system generates both a literal copy (exact replication of the original path) and adjusted versions that reinterpret the path in the new context (e.g., "Urban Busker / Gouache," "Night Busker / Vivid Watercolor").
    (c) Reusing Whole Projects. Users can import a complete project graph, either their own or from another designer. Imported projects can be brought in literally (preserving all blocks exactly) or adaptively (automatically adjusting properties to fit the new project's topic). For example, importing an "Urban music festival mascot" project into a new "Kids' coding platform mascot" project transforms elements such as "Street Musician" into "Programmer" and adapts image styles to more child-friendly aesthetics.}
    \label{fig:feature3}
\end{figure*}

To support efficient and flexible exploration, \sysname{} provides multiple features for reusing earlier divergent intents along two complementary axes: the unit of reuse (block, path, or project), and the level of adaptation (literal vs. adaptive).

\paragraph{\textbf{Block-level Reuse}.}
At the most granular level, users can reuse individual \textit{Exploration Blocks}, which corresponds to each exploration instance. All previously created blocks across projects are stored in the left sidebar (Figure~\ref{fig:feature3} (a)). As users explore, \sysname{} uses GPT-4o-mini to generate a graph in the left panel that visualizes how their exploration directions evolve for each property. Selecting a block from this graph places a new copy onto the canvas, which users can then connect to new exploration paths. \sysname{} supports both literal reuse (reusing the block as-is with the same parameters and suggestions), and adaptive reuse (generating adjusted variants of the block based on the current exploration context). All adaptations (at the block, path, and project levels) are generated using GPT-4o with the same prompts.

\begin{block}
During the exploration, \username{} explores a new background setting, `Deep sea stage', and wants to reuse a previously explored `Watercolor' image style on it. He selects the saved block in the sidebar and connects the copied block to the new setting. \sysname{} then presents two options: a literal version reproducing the earlier style suggestions, and an adaptive version retaining the watercolor aesthetic while tailoring it to the `deep sea' context.
\end{block}

\paragraph{\textbf{Path-level Reuse}.}
Users can also reuse an entire sequence of previously crafted blocks by copying and pasting an exploration path (Figure~\ref{fig:feature3} (b)). By choosing a path to reuse and a point on the canvas to paste, two options are given: (1) a literal copy that replicates all blocks as-is, or (2) a context-adaptive copy that adjusts block directions and typicality levels based on the context of the target point.
Once pasted, the system creates blocks with appropriate parameters and automatically continues the exploration, which enables users to efficiently branch off from a new starting point while preserving a familiar exploration trajectory.

\begin{block}
Later, \username{} recalls a successful exploration path about `Street Musician' character paired with a `Watercolor' style. Eager to reuse it, he selects that path and pastes it on the new `Skyscraper' background setting. \sysname{} offers literal copy and three different context-adaptive variants. In one adaptive version, the character is reimagined as a `Night Busker', and the style shifts to `Gouache' to better align with the current `Skyscraper' theme. \username{} selects this version, quickly branching into a new yet familiar sequence of exploration.
\end{block}

\paragraph{\textbf{Project-level Reuse}.}
At the broadest unit, \sysname{} allows users to import an entire project, including all paths and blocks. Users can reuse their own or other users’ projects on similar topics (Figure~\ref{fig:feature3} (c)).
As with block- and path-level reuse, \sysname{} offers literal reuse (preserving all blocks as-is) and adaptive reuse, which adjusts the parameters of each block to better align with the new project topic.
Imported projects appear on the canvas, which users can then extend, branch from, or merge with their current exploration.

\begin{block}
After working on the mascot design for a `Music Festival', \username{} starts a new mascot design for `Kids' coding platform'. Rather than starting from scratch, he imports the earlier project about a music festival mascot. \sysname{} offers literal and adaptive reuse options; in one adaptive version, \username{} found that the `Street Musician' entity is reimagined into `Digital Nomad', and `Watercolor' style shifts to `Illustration' to better suit a playful, child-friendly theme. By choosing this, \username{} can quickly bootstrap his new exploration.
\username{} can also browse projects created by other users working on similar topics. While looking at others' explorations, he discovers an interesting branch that combines an `Animal Character' with `Sticker' style. He imports this project and borrows that segment into his own flow. This allows him to blend familiar ideas with fresh perspectives from others’ explorations.
\end{block}

\subsection{Technical Details \& Pipeline Overview}

\sysname{} was built as a web-based system using ReactJS\footnote{https://react.dev/} and React Flow\footnote{https://reactflow.dev/}. \sysname{} includes a pipeline that operationalizes divergent intent by translating user-specified intent parameters into system-generated suggestions. The full prompts used for each step of this pipeline are provided in the supplementary material.
% \sysname{} includes two pipelines that operationalize divergent intent for interaction: (1) translating user-specified intent parameters into system-generated suggestions, and (2) generating images that incorporate prior exploration history.
% All the prompts used in \sysname{} are in Supplementary Material.

\subsubsection{Divergent Intent into System-Generated Suggestions}\label{sec:tech1}

A key mechanism of \sysname{} enables users to explicitly define the design space by controlling three parameters of divergent intent: property, direction, and range.
We designed and evaluated one concrete instantiation of this concept through our pipeline that translates users' input on visual property, direction keyword, and typicality range into controlled visual or textual suggestions.

\paragraph{\textbf{Pipeline Design}.}
\sysname{} first generates a diverse pool of candidate suggestions using a multi-step prompting strategy. Given the input property and direction, GPT-4o generates ten alternative directions spanning from literal extensions of the input to more distant interpretations. For each alternative direction, the system produces multiple suggestions, yielding a total of 100 text-based or 50 image-based candidates. This pool serves as the raw material for the subsequent algorithmic filtering and selection steps.

Next, to follow the user-specified `range' (typicality level), the system computationally filters the candidate pool.
We define typicality as how easily a suggestion can be mentally evoked from the user's input direction (i.e., how commonly the two co-occur conceptually or linguistically). To operationalize this, we measure co-occurrence-based similarity between the input direction and each candidate: for text, using GloVe~\cite{gan2022semglove, pennington2014glove}; and for images, using the CLIP score~\cite{radford2021learning}. By ranking and filtering candidates based on these calculated metrics, the system ensures that the output aligns with the user's intended magnitude of divergence.

Finally, to support iterative refinement of `Direction', the system employs K-means clustering to structure them into groups. We cluster the candidates using Sentence-BERT~\cite{reimers2019sentence} (text) or ViT-B/32 CLIP embeddings~\cite{radford2021learning} (image) and present the centroids of the four most distinct clusters first. When a user selects or refines one of these suggestions, the system utilizes the embedding of the chosen centroid as the new semantic anchor, dynamically shifting the direction.

\paragraph{\textbf{Pipeline Evaluation}.}
We evaluated the pipeline to examine whether its typicality-based filtering aligns with human judgment. We constructed an evaluation set using five \((Property, Direction)\) pairs from the formative study (3 text-based, 2 image-based). Using these pairs, the pipeline generated 100 candidates per text property and 50 per image property, totaling 400 suggestions. After that, we randomly selected 100 suggestions (60 text-based, 40 image-based) as a final evaluation set.
We recruited 3 raters with prior design experience and asked them to label typicality levels for each system-generated suggestion given the input direction and property. Inter-rater reliability (Krippendorff's Alpha) was 0.75 for text-based and 0.69 for image-based suggestions, indicating substantial agreement. Final labels were established by averaging their ratings.

We then examined the alignment between human-labeled typicality levels and the pipeline’s typicality ordering. For text-based suggestions, the system's scores showed moderately strong alignment with human annotations (Pearson \(r\)=0.68, Spearman \(\rho\)=0.67). For image-based suggestions, correlations were moderately aligned (Pearson \(r\)=0.59, Spearman \(\rho\)=0.58). Together, these results indicate that the pipeline produces suggestions whose typicality ordering broadly corresponds to human perception.

We note that the metrics that we chose serve as proxies for typicality: GloVe 
captures thematic co-occurrence rather than categorical typicality (i.e., how representative an instance is of its category~\cite{rosch1973on}), and CLIP measures image-text correspondence rather than concept-level co-occurrence. While our pipeline evaluation showed reasonable alignment with human judgments, these remain approximations of the underlying theoretical construct.

%% file: Sections/05-Evaluation.tex
\section{Evaluation}

To evaluate \sysname{} and understand its role in supporting design exploration, we conducted two complementary studies. First, we conducted a within-subjects comparative study to validate whether \sysname{} effectively supports divergent intent expression and reuse, ultimately leading to more divergent design exploration than the baseline. This study addressed the following research questions:

\begin{itemize}
    \item RQ1. How does \sysname{} support users in expressing and reusing their divergent intents?
    \item RQ2. How does the use of \sysname{} influence users' exploration process and patterns?
    \item RQ3. How does \sysname{} impact the ideational fluency and diversity of users' exploration?
\end{itemize}

We then conducted a three-day longitudinal study to more closely examine \sysname{}'s intent-reuse features.
While the formative study provided a foundation for the design of a divergent intent expression, it revealed broad possibilities for designing reuse features, leading to multiple intent-reusing features in \sysname{}. Therefore, we investigated how each feature is used in practice --- when and why designers engage with it, and what purposes it serves in exploration. From these observations, we derived design implications for future intent-reusing tools. This study addressed the following research questions:

\begin{itemize}
    \item RQ4. When and why do designers engage with reuse features with different units and mechanisms in practice?
    \item RQ5. How should future intent-reuse tools be designed to better support creative exploration?
\end{itemize}

%% file: Sections/06-Study1.tex
\section{Study 1: Comparative Study}
We conducted a within-subjects comparative study with 12 participants to assess how well \sysname{} supports the expression and reuse of divergent intents. Additionally, we analyzed its impact on users' exploration patterns and the ideational fluency and diversity of their exploration outcomes to check if it helped divergent exploration.

\subsection{Study Design}

While divergent and convergent intents are interleaved in the design process, they serve different cognitive goals and are supported by different mechanisms. To examine whether and how \sysname{}'s intent expression and reuse benefits divergent exploration, we intentionally focused more on the effects related to divergent intent in this study, rather than evaluating overall creative quality. We discussed the role of convergent processes in overall creative outcomes in Section~\ref{sec:convergence}.

\subsubsection{Participants}

We recruited 12 participants (5 females; $M$=25, $SD$=1.71) through an online recruitment post. All had a design background through education or professional experience. Six were graduate students in design, three were final-year undergraduates in design, two held design degrees and were working, and one had no formal degree but worked as a graphic designer. Details are provided in Appendix~\ref{appendix:studyparticipant}. The study lasted about 1.5 hours, and participants were compensated 50,000 KRW (approx \$34 USD).

\subsubsection{Study Procedure}
Participants performed two rounds of design exploration tasks, each using either \sysname{} or the baseline system. The task involved generating diverse design concepts for one of two design topics: \textit{``An illustration for a book cover of a detective novel set in a future city''} or \textit{``A mascot for a summer night urban film festival.''} These topics are selected as they do not have a typical visual representation and involve diverse design elements to explore, so that participants can explore divergently. The order of topics and systems was counterbalanced across participants.

Each round began with a 10-minute tutorial on the given system, during which participants practiced with a sample design topic to familiarize themselves with the interface. They then performed the exploration task for 20 minutes while thinking aloud about how their ideas developed. After each round, participants completed a post-task survey. Following both sessions, we conducted a 15-minute semi-structured interview. The full set of interview questions is in Appendix~\ref{appendix:interviewquestions}.

\subsubsection{Systems}

We designed a baseline system as a generative AI–based design exploration tool with the same canvas-based interface as \sysname{}, but with only conventional mechanisms for intent expression and reuse. A screenshot is shown in Figure~\ref{fig:screenshot-baseline} in the Appendix.
In the baseline, users enter a free-form text prompt describing their desired exploration, and the system generates four images using DALL·E 3, the same image generation model as \sysname{}. Like \sysname{}, the baseline supports non-linear exploration through branching on the canvas. However, its reuse mechanism simply copies the text prompt from the preceding block when a new block is added and linked, without additional structuring or contextual adaptation.
To offset the absence of \sysname{}’s LLM-based pipelines for intent expression and reuse (e.g., interpreting divergent intent or context-aware block copying), participants in the baseline condition were given access to ChatGPT\footnote{https://chat.openai.com/}.
Critically, the only difference between the two conditions was whether divergent intent was expressed through structured Exploration Blocks or a free-form text prompt, allowing us to isolate the effects of \sysname{}'s intent-expression and reuse mechanisms.

We excluded the project-level reuse feature from \sysname{} in this study, as it cannot be meaningfully evaluated within a 20-minute session using a single exploration topic. Its use was examined in the longitudinal study (Section~\ref{sec:study2}) instead.

\subsubsection{Measures}

We collected both qualitative and quantitative measures to assess how the system supported the divergence of early-stage design exploration.
The post-task survey included 7-point Likert ratings on system support for expression and reuse of divergent intent, perceived efficiency and diversity of idea exploration, and satisfaction with exploration outcomes. Additionally, the survey also included the Creativity Support Index (CSI)~\cite{csi} and NASA-TLX~\cite{HART1988139}.
We also analyzed timestamped behavioral logs of actions (e.g., block creation, image generation, feature use) to derive quantitative measures, including the number and rate of generated images and the diversity of outputs. Finally, we analyzed the structural characteristics of the exploration session using linkography~\cite{goldschmidt2014linkography}, capturing branching patterns, convergence, reuse, and connectivity between ideas.
We used the Wilcoxon signed-rank test for all Likert-scale survey responses. For log data, we first ran a Shapiro-Wilk test for normality and applied either a paired t-test or a Wilcoxon test accordingly.

\subsection{Results}

Our findings indicate that \sysname{} significantly enhanced users' ability to express and reuse their divergent intent. It also supported more interconnected and structured patterns of idea exploration, resulting in greater ideational fluency and a broader diversity of concepts explored. We present our results in alignment with our three research questions (RQ1-3).

\subsubsection{Support for Expressing and Reusing Divergent Intents}
To answer RQ1, we examined survey responses and log data to evaluate how effectively participants could express and reuse their divergent intents.

\paragraph{Expressing Intent}

\begin{figure}[h]
    \centering
    \includegraphics[width=\linewidth]{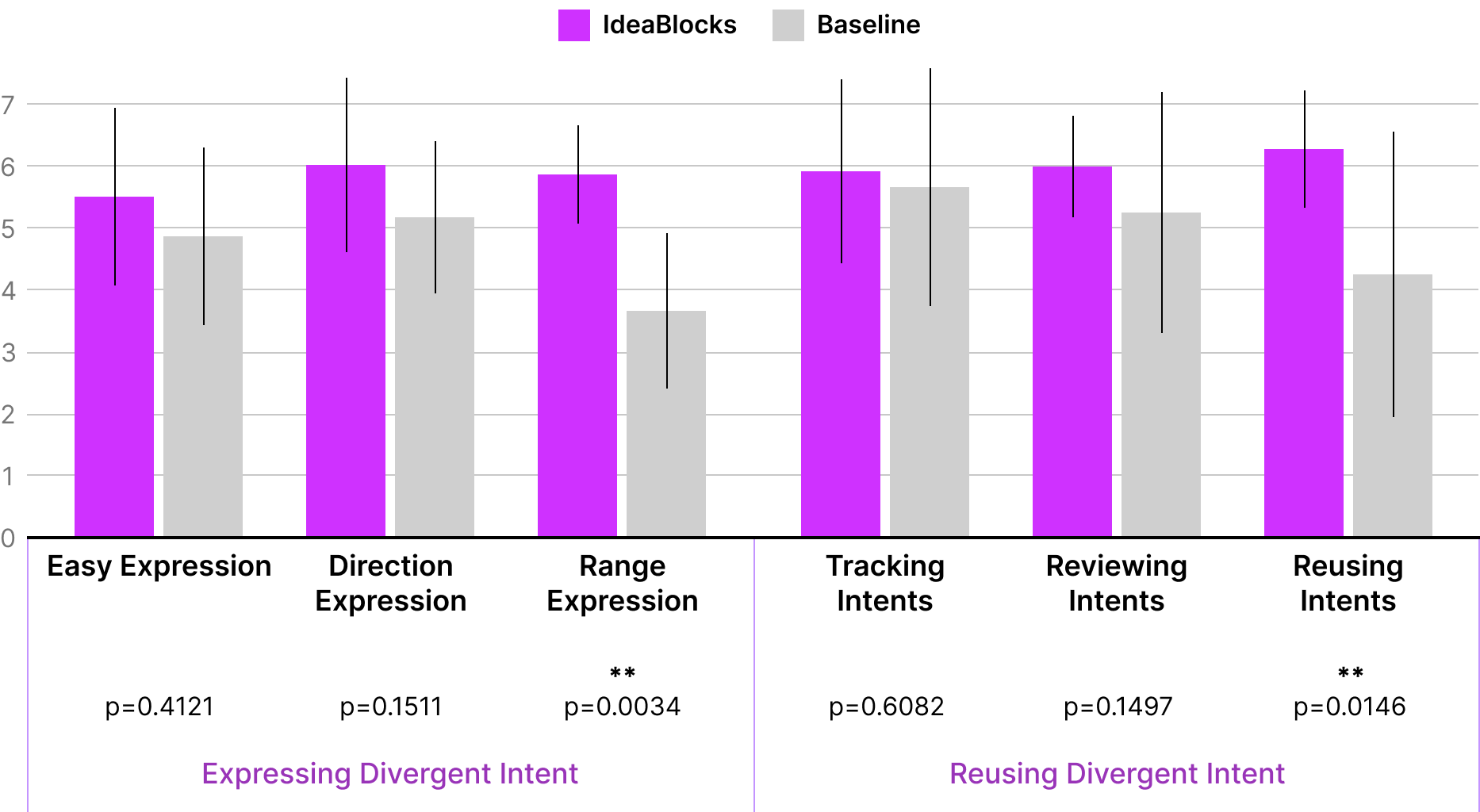}
    \caption{Survey responses on users' perceived support for intent expression (left) and intent reuse (right), with 95\% confidence intervals. Participants rated \sysname{} significantly higher in their ability to express a range of intent and to reuse previously expressed intent.}
    \Description{
    Bar chart comparing user ratings between \sysname{} and the baseline across six aspects: three on intent expression (easy expression, direction expression, range expression) and three on intent reuse (tracking, reviewing, and reusing intents). \sysname{} consistently received higher ratings, with statistically significant differences observed in range expression (p=0.0034) and reusing intents (p=0.0146). Error bars represent 95\% confidence intervals.
    }
    \label{fig:survey-rq1}
\end{figure}

As shown in Figure~\ref{fig:survey-rq1}, participants rated \sysname{} significantly more helpful in expressing the \textit{range} of their divergent intent ($M$=5.83, $SD$=0.80) compared to the baseline ($M$=3.67, $SD$=1.25 / $p$=0.0034, $W$=4.5). Participants largely attributed this advantage to the system's \textit{typicality} slider, which allowed them to control how far the exploration should deviate, thereby better articulating the intended scope of their exploration.
Although \sysname{} also received higher average scores in supporting the expression of \textit{direction}, the difference was not statistically significant. This is likely because participants could still partially control directional intent through the baseline's text prompt.
% However, interview responses revealed that \sysname{}'s feature to iterate on suggestions helps them steer the directions based on their evolving ideas. P10 stated \textit{``It was nice that the system showed the examples (...) It helped me better understand what kind of exploration it was doing and control it.''}

This difference in perceived expressiveness was also reflected in participants' behaviors. Participants with \sysname{} created more input blocks ($M$=23.25, $SD$=7.97) compared to the baseline ($M$=13.17, $SD$=2.66 / $p$=0.0010, $t$ = -4.41).
% The higher volume of user-generated inputs suggests that \sysname{} reduced the interaction friction of articulating divergent intent. As noted in interviews, by providing a structural scaffold for intents that were previously unstructured and ambiguous in text prompts, the system enabled users to express their goals more frequently and fluidly.
During the interview, participants described how \sysname{} made intent expression easier. Rather than formulating a fully specified prompt upfront, they treated intent expression as an incremental and adjustable process. By separating property, direction, and range, \sysname{} allowed participants to externalize partial intent even when their goals were still unclear, and to refine it through interaction. As a result, participants perceived intent expression as a low-commitment activity, which encouraged more frequent articulation throughout the exploration session.

\paragraph{Reusing Intent}

As illustrated in Figure~\ref{fig:survey-rq1}, participants with \sysname{} reported greater effectiveness in reusing prior intents ($M$=6.25, $SD$=1.01) than the baseline ($M$=4.25, $SD$=2.31 / $p$=0.0146, $W$=2).

During the 20-minute sessions, participants using \sysname{} engaged in intent-reusing actions (copying blocks or paths) an average of 5.08 times ($SD$=4.36). Out of the 61 total instances of intent reuse, 41 targeted visual properties (e.g., image style, color palette). According to interview responses, participants found reuse particularly useful for such visual aspects, as these preferences tended to remain consistent across different ideas. In contrast, text-based properties (e.g., character entities) were seen as more context-dependent and were frequently changed between ideas. This contrast also appears in the longitudinal study findings in Section~\ref{sec:blockreuse}.

Ratings for tracking and reviewing intent did not differ significantly between conditions, as the interview results showed that these aspects were largely supported by the non-linear canvas interface that both baseline and \sysname{} shared.
However, their reasons for doing so differed between conditions. P11 explained, \textit{``In the baseline, I reviewed my exploration flow to assess the outputs and determine how to fix my next prompt. In \sysname{}, I do that to see which results were good to copy them directly.''} This suggests that \sysname{} also shaped how participants reviewed prior work with a clearer purpose and actionable follow-up.

\begin{table*}[]
\begin{tabular}{rrcccccc}
\hline
                                                                                &                            & \multicolumn{2}{c}{\sysname{}} & \multicolumn{2}{c}{Baseline} & \multicolumn{2}{c}{Statistics} \\ \cline{3-8} 
                                                                                &                            & $M$                 & $SD$                  & $M$          & $SD$          & $p$                   & Sig.     \\ \hline
\multirow{3}{*}{\begin{tabular}[c]{@{}r@{}}Linkography\\ Analysis\end{tabular}} & Avg. link distance         & 3.512                & 1.441                & 2.459         & 0.913        & \textbf{0.0471}     & \textasteriskcentered{}        \\ \cline{2-8} 
                                                                                & \# of connected components & 1.917                & 0.793                & 4.333         & 1.497        & \textbf{0.0031}     & \textasteriskcentered{}\textasteriskcentered{}       \\ \cline{2-8} 
                                                                                & Link entropy               & 1.152                & 0.524                & 0.537         & 0.589        & \textbf{0.0409}     & \textasteriskcentered{}        \\ \hline
\multirow{2}{*}{Ideational Fluency}                                                     & \# of input block          & 23.25               & 7.97                & 13.17            & 2.66        & \textbf{0.0010}     & \textasteriskcentered{}\textasteriskcentered{}      \\ \cline{2-8} 
                                                                                & \# of generated images      & 123.42                & 41.66                 & 58         & 13.27         & \textbf{0.0005}     & \textasteriskcentered{}\textasteriskcentered{}\textasteriskcentered{}       \\ \hline
Diversity                                                                       & Max cosine distance        & 0.6384               & 0.0505               & 0.5675        & 0.0897       & \textbf{0.0457}     & \textasteriskcentered{}        \\ \hline
\end{tabular}
\caption{Summary of quantitative results across RQ1–RQ3. Metrics include: (1) linkography-based structural measures (e.g., link distance, entropy), (2) behavioral indicators of ideational fluency (e.g., number of input blocks, generated images), and (3) diversity measures using CLIP-based cosine distance. (\textasteriskcentered{}: p < .050, \textasteriskcentered{}\textasteriskcentered{}: p < .010, \textasteriskcentered{}\textasteriskcentered{}\textasteriskcentered{}: p < .001)}
\Description{This table presents the quantitative results comparing \sysname{} with a baseline system across three dimensions: linkography-based structural measures (RQ2), ideational fluency (RQ1, RQ3), and output diversity (RQ3). For linkography analysis, \sysname{} achieved a higher average link distance, fewer connected components, and greater link entropy, suggesting a more structured and interconnected exploration. Regarding fluency, users of \sysname{} created more input blocks and generated more images, indicating increased ideation throughput. For diversity, the maximum CLIP-based cosine distance was higher in the \sysname{} condition, reflecting broader visual variation in the results.}
\label{tab:results}
\end{table*}

\subsubsection{Exploration Patterns and Process}

To address RQ2, we examined participants' exploration patterns using linkography~\cite{goldschmidt2014linkography}, which analyzes the progression of ideas by representing each idea as a node and connections (e.g., references, derivations) as directed links.
In our study, we constructed a linkograph for each session, where each node corresponds to an input block. Links captured (1) parent–child relationships formed through the non-linear canvas in both \sysname{} and the baseline, (2) intent-reuse actions unique to \sysname{}, and (3) explicit references to earlier ideas made during think-aloud.
Example linkographs are provided in Figure~\ref{fig:linkograph-example} in the Appendix.

\paragraph{Diverse and Interconnected Idea Development}

As shown in Table~\ref{tab:results}, \sysname{} sessions exhibited significantly higher link entropy ($M$=1.15, $SD$=0.52) than the baseline ($M$=0.54, $SD$=0.59 / $p$=0.0409), indicating that participants were more likely to pursue multiple parallel conceptual directions rather than a narrow or linear trajectory~\cite{goldschmidt2014linkography}.
In addition, ideas generated with \sysname{} were connected across longer temporal and conceptual spans, as reflected by a greater average link distance ($M$=3.51, $SD$=1.44) compared to the baseline ($M$=2.46, $SD$=0.91 / $p$=0.0471). \sysname{} sessions also exhibited significantly fewer connected components ($M$=1.92, $SD$=0.79) than the baseline ($M$=4.33, $SD$=1.50 / $p$=0.0031), suggesting a more cohesive exploration structure overall.

Rather than reusing prior ideas to converge quickly or finalize results, participants described reusing earlier intents to keep diverse alternative directions available, which they referred to as ``deferring commitment'' to any single direction. In this sense, the increased connectivity and long-range linking observed with \sysname{} reflect deliberate maintenance of multiple viable directions, supporting sustained and exploratory navigation of the design space. Some participants also reflected on recognizing recurring exploration strategies through reuse, a pattern that we further examine in the longitudinal study in Section~\ref{sec:study2}.

% As shown in Table~\ref{tab:results}, \sysname{} sessions exhibited significantly higher link entropy ($M$=1.15, $SD$=0.52) than the baseline ($M$=0.54, $SD$=0.59 / $p$=0.0409, $W$=10). This indicates that participants using \sysname{} developed ideas in a more diverse and interconnected manner compared to the baseline. Instead of following a narrow or linear trajectory, they pursued multiple parallel conceptual directions, reflecting more divergent exploration patterns~\cite{goldschmidt2014linkography}.

% \paragraph{Long-Range and Cohesive Idea Development}

% The average link distance (the temporal span between connected ideas) was significantly greater with \sysname{} ($M$=3.51, $SD$=1.44) compared to the baseline ($M$=2.46, $SD$=0.91 / $p$=0.0471, $t$ = -2.2346), indicating that participants were more likely to connect ideas across longer time spans or conceptual gaps, rather than relying only on their most recent ideas.
% \sysname{} sessions also had significantly fewer connected components ($M$=1.92, $SD$=0.79) than the baseline ($M$=4.33, $SD$=1.50 / $p$=0.0031, $W$=0), suggesting that their ideas were organized into a more cohesive exploration structure.
% Together, these patterns indicate a more iterative and reflective exploration process, in which participants revisited, extended, and integrated previous ideas rather than producing isolated clusters.

\subsubsection{Ideational Fluency and Flexibility of Exploration}

Creativity theory~\cite{guilford1967nature, torrance1966torrance} suggests that divergent thinking is driven by \textit{ideational fluency} (the ability to generate many responses) and \textit{flexibility} (the ability to explore diverse categories). Because \sysname{} targets early-stage expansion of the design space rather than refinement or evaluation, we focus on these two dimensions to assess whether it supports divergent exploration.

\begin{figure*}
    \centering
    \includegraphics[width=0.85\linewidth]{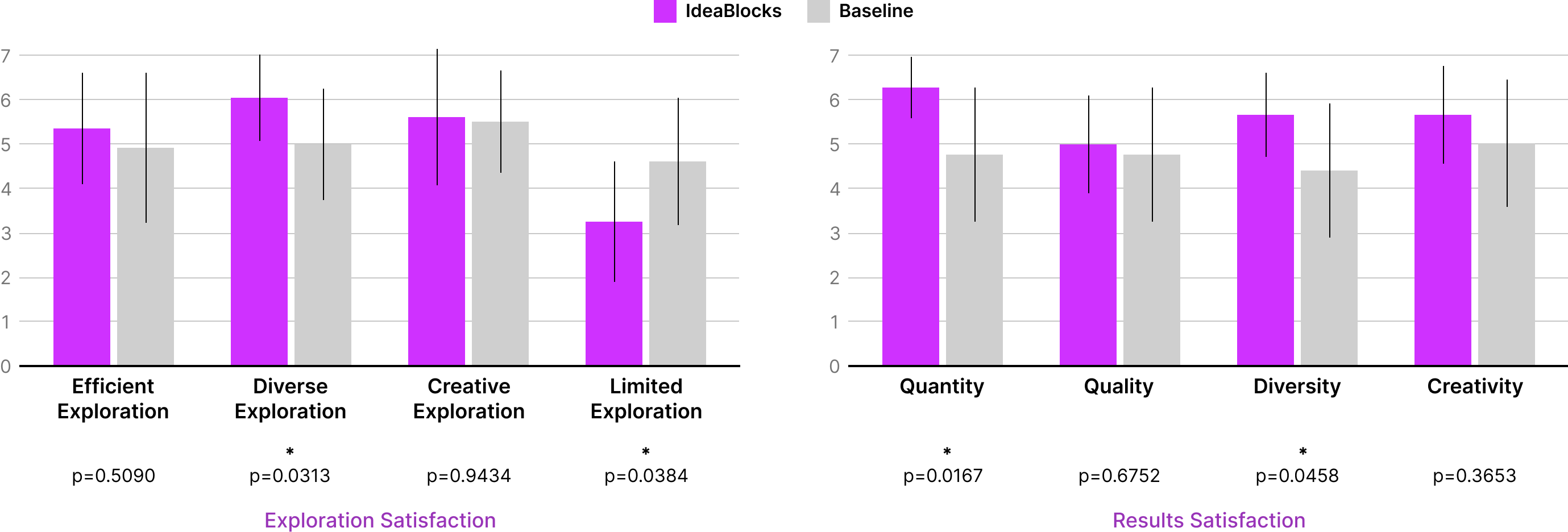}
    \caption{
    Survey responses on users' perceived satisfaction for the exploration process (top) and satisfaction with exploration results (right), with 95\% confidence intervals. Participants rated \sysname{} significantly higher in supporting diverse exploration, reducing limited exploration (reversed item), and improving satisfaction with both the quantity and diversity of outcomes.
    }
    \Description{
    Two grouped bar charts comparing user ratings between \sysname{} and the baseline. The top chart shows satisfaction with the exploration process, where \sysname{} received significantly higher ratings for diverse exploration and lower ratings for limited exploration. The bottom chart presents satisfaction with outcomes, where participants rated \sysname{} significantly higher for quantity and diversity of results. Error bars indicate 95\% confidence intervals.
    }
    \label{fig:survey-rq3}
\end{figure*}

\paragraph{Ideational Fluency}

\begin{figure}[h]
    \centering
    \includegraphics[width=\linewidth]{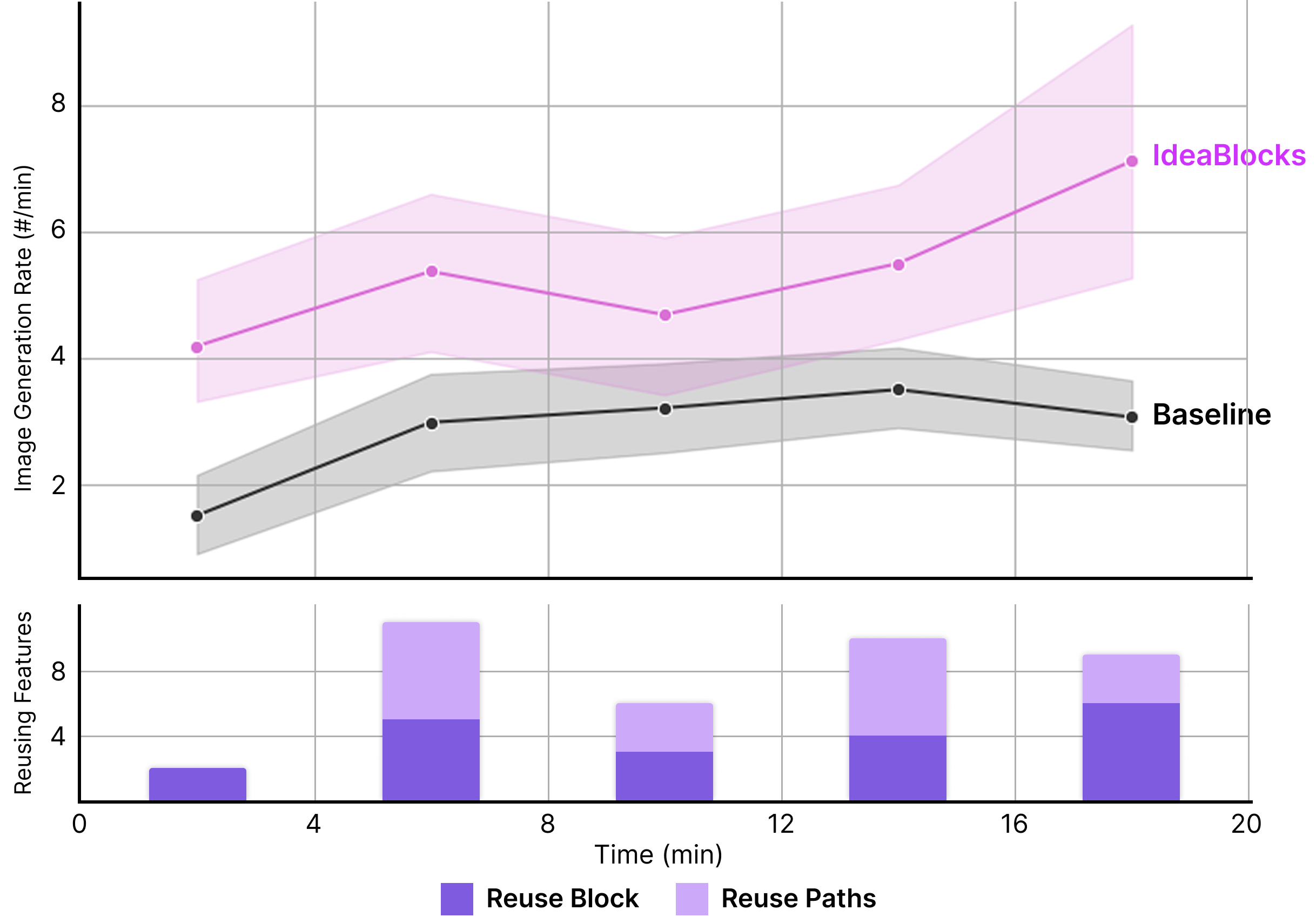}
    \caption{
    (Top) Image generation rate over time with 95\% confidence intervals, grouped by 4-minute segments. \sysname{} users showed a steady increase, while the baseline plateaued mid-session. (Bottom) Frequency of reuse feature usage, split by block-level (\textit{Copy Blocks}) and path-level (\textit{Copy Paths}). The rising use of reuse features aligns with the acceleration in generation speed, suggesting their role in sustaining exploration momentum.
    }
    \Description{
    The figure contains two aligned plots. The top plot is a line chart showing the image generation rate over time. The x-axis represents five time segments (0–4, 4–8, ..., 16–20 minutes), and the y-axis represents images generated per minute. Two lines are plotted: a pink line for \sysname{} showing a steady upward trend, and a black line for the baseline, which increases initially but flattens after 8 minutes. The bottom plot is a stacked bar chart aligned with the same time segments. Each bar represents the total number of reuse feature uses in \sysname{}, broken down into two categories: darker purple bars for \textit{Copy Blocks} and lighter purple overlays for \textit{Copy Paths}. The chart shows that reuse interactions increase over time, with a noticeable spike in the 16–20 minute segment. This corresponds to the rise in the generation rate in the top plot.
    }
    \label{fig:image-creation-speed}
\end{figure}

% In this work, we interpret ideational fluency not as indiscriminate generation, but as the ability to sustain exploration by building upon previously articulated intent rather than repeatedly starting from scratch.
As shown in Table~\ref{tab:results}, participants demonstrated significantly higher ideational fluency when using \sysname{}, generating more than twice as many images during the 20-minute session ($M$=123.42, $SD$=41.66) compared to the baseline ($M$=58.00, $SD$=13.27 / $p$=0.0005). Figure~\ref{fig:image-creation-speed} shows that participants using \sysname{} also exhibited increasing image generation rates over time, whereas the baseline condition showed no such acceleration.
The interview suggested that this increase was driven by their ability to build upon existing intentions rather than repeatedly starting from scratch. Reuse features allowed participants to preserve previously articulated intent while exploring variations, reducing the effort required to explore new ideas. As a result, participants were able to sustain exploration momentum over time, which aligns with the observed acceleration in image generation during later stages of the session.

% As in Table~\ref{tab:results}, participants demonstrated higher \textbf{ideational fluency} when using \sysname{}, generating more than double the number of images during the 20-minute session ($M$=123.42, $SD$=41.66) than in baseline ($M$=58.00, $SD$=13.27 / $p$=0.0005, $W$=0).
% Furthermore, Figure~\ref{fig:image-creation-speed} shows that \sysname{} enabled participants to accelerate their image generation over time. 
% Between the final two segments of the session (12–16 and 16–20 minutes), their rate increased by 1.62 images/min, while the baseline condition showed a slight decrease of –0.44 ($p$=0.0491, $t$ = –2.21).
% Figure~\ref{fig:image-creation-speed} shows that accelerations in image generation rate aligned with peaks in the use of \sysname{}'s reuse features. Participants frequently reused prior intents (via block and path reuse) during key segments of the session (4–8 and 12–20 minutes), coinciding with noticeable increases in generation speed. Interviews revealed that these reuse features enabled participants to build upon existing directions rather than starting from scratch, supporting faster and more confident exploration.

Participants reported significantly higher satisfaction with the quantity of ideas generated using \sysname{} ($M$=6.25, $SD$=0.72) compared to the baseline ($M$=4.75, $SD$=1.53 / $p$=0.016, $W$=2.5), but their ratings of exploration efficiency ($M$=5.33, $SD$=1.25) did not differ significantly from the baseline ($M$=4.92, $SD$=1.66; $p$=0.509, $W$=17). Interview responses suggest that participants interpreted efficiency differently: some associated it with how quickly they could converge on a final outcome (efficiency in convergence), while others focused on sustaining exploration (fluency in divergence). From this perspective, the observed increase in ideational fluency reflects a form of efficiency specific to divergent exploration, rather than faster convergence toward a single result.

% These behavioral measures of fluency were consistent with survey participants' perceived satisfaction regarding the \textit{quantity} of their ideas ($M$=6.25, $SD$=0.72) compared to the baseline ($M$=4.75, $SD$=1.53 / $p$=0.016, $W$=2.5).
% However, when asked specifically whether the system helped them explore \textit{efficiently}, ratings for \sysname{} ($M$=5.33, $SD$=1.25) were not significantly different from the baseline ($M$=4.92, $SD$=1.66; $p$=0.509, $W$=17). Interview responses suggested that this discrepancy stems from the nuanced difference between \textit{fluency} and \textit{efficiency} in design: while \sysname{} supported the \textit{divergent} goal of generating many options (fluency), some participants interpreted ``efficiency'' through a \textit{convergent} lens, focusing on how quickly they could narrow down to a final satisfactory result rather than how many ideas they could explore.

\paragraph{Flexibility of the Results}
Beyond the volume of ideas generated, flexibility captures whether exploration spans diverse conceptual directions rather than variations within a narrow scope. To assess this aspect, we analyzed the diversity of the generated outcomes at the image level.
Using CLIP-based embeddings, we measured the maximum pairwise distance among images generated in each session as an indicator of visual diversity. As shown in Table~\ref{tab:results}, \sysname{} led to more visually diverse outcomes ($M$=0.6384, $SD$=0.0505) than the baseline ($M$=0.5675, $SD$=0.0897 / $p$=0.0457, $t$ = -2.2529), suggesting that \sysname{} supported exploring a broader range of ideas within the given time.
In the survey, participants also reported significantly higher satisfaction with the diversity of their outcomes when using \sysname{} ($M$=5.67, $SD$=0.94) than with baseline ($M$=4.42, $SD$=1.50 / $p$=0.043, $W$=6).

\paragraph{Perceived Creativity}
Despite the positive results for divergent exploration, ratings of how much the system helped them explore creatively were similar (Baseline: $M$=5.50, $SD$=1.12 / \sysname{}: $M$=5.58, $SD$=1.55 / $p$=0.591, $W$=17.5), and satisfaction with the creativity of the final results did not differ significantly (Baseline: $M$=5.00, $SD$=1.47 / \sysname{}: $M$=5.67, $SD$=1.11 / $p$=0.278, $W$=15). Similarly, CSI scores showed no statistically significant differences between conditions. Together, these findings suggest that while \sysname{} effectively supports divergent exploration, the overall creativity requires complementary convergent processes as well, such as selection, refinement, and evaluation, which were intentionally held constant in this study. We discussed this complementary relationship between these two and suggested future directions in Section~\ref{sec:convergence}.

% \paragraph{Task Load}
% The NASA-TLX results showed no significant difference across any of the six dimensions, indicating that \sysname{}’s additional functionalities did not increase the perceived task load. Full results are shown in Appendix~\ref{appendix:fullsurvey}.

%% file: Sections/07-Study2.tex
\section{Study 2: Longitudinal Study}\label{sec:study2}
We conducted a three-day longitudinal study with six participants to understand when and how people use different intent-reuse features, and how the system design could better support them. Based on the findings, we suggest design implications for future divergent intent reuse systems.

\subsection{Study Design}

\subsubsection{Participants}
We recruited 6 participants (3 females; $M$=25, $SD$=1.71) through an online recruitment post.
Four were professional designers with prior work experience: two in graphic design (2 and 3 years), and two in UI design (6 months and 2 years). The other two were design students, one graduate and one undergraduate. Participant details are provided in Appendix~\ref{appendix:studyparticipant}. One participant (D3) had also taken part in the comparative study (P9). Each participant was compensated 100,000 KRW (approx \$72 USD).

\subsubsection{Study Procedure}
After the 30-minute system tutorial session, each participant used \sysname{} over three days, completing two 25-minute design exploration tasks per day (one mascot design and one book cover illustration task; $\approx$50 minutes of use per day). Throughout the study, \sysname{} collected usage logs, including timestamped actions such as the use of different intent-reuse features. After completing all tasks, we conducted a 30-minute semi-structured interview with each participant. The full interview protocol is provided in Appendix~\ref{appendix:interviewquestions2}.

\subsection{Results}

\subsubsection{Block-level Reuse: Visual Assets as Designers' Palettes}\label{sec:blockreuse}

During the study, participants reused single blocks, averaging 2.19 copy actions (\(SD\)=2.36) per 25-minute project. For each reuse, users could choose literal or adaptive pasting. Interestingly, participants strongly preferred literal reuse for visual properties such as image styles or color palettes (35 literal vs. 9 adaptive). Participants explained that these visual assets reflected their global preferences and were reused across projects. In contrast, for semantic properties such as characters’ entities or poses, reuse was more evenly split (18 literal vs. 17 adaptive), suggesting that these aspects are more context-dependent and often require adaptation.

Several participants explicitly described block-level reuse as a way to build and reuse a ``personal palette.''
D1 explained, \textit{``Poses are highly dependent on context, so they are harder to reuse as-is, but image styles and color palettes are something I often bring back. They are closely tied to my personal taste.''} D4 made an analogy to design tools, explaining that \textit{``just like storing and reusing color swatches, I treated styles and colors as palettes to apply across projects.''}

Even for non-visual semantic properties, participants preferred literal reuse when they had strong familiarity with a specific property type. For example, D5, who regularly designs YouTube thumbnails, explained: `\textit{`For background settings, I usually have a very clear sense of what I want, probably because I work with backgrounds a lot, so I already had plenty of ideas to reuse without any changes.''}

\begin{blockBig}
\textbf{Design Implication:} Systems should differentiate reuse mechanisms based on both property type and user expertise. Visual properties, or properties where users have strong familiarity, may benefit from literal reuse as cross-project palettes that reinforce identity and consistency. In contrast, semantic properties may be better supported through adaptive reuse, enabling greater contextual flexibility.
\end{blockBig}

\subsubsection{Path-level Reuse: Template-like Scaffolding}

Participants also engaged with copying and pasting entire exploration paths, and we observed that this enabled users to reuse not just multiple blocks but the \textbf{structure of the exploration process} itself.
They often described their exploration paths as recurring sequences of properties (e.g., entity $\rightarrow$ style $\rightarrow$ pose for D1) that functioned as reusable scaffolds. In interviews, they mentioned that, over multiple days of using \sysname{}, they began to recognize their own recurring strategies and increasingly relied on path reuse to reapply them quickly.
This shift was reflected in the logs: use of path reuse grew steadily, from an average of 2.10 uses early on Day 1 to 2.90 on Day 2 and 3.30 on Day 3. Block- and project-level reuse showed no such upward trend.

D2 imagined formalizing such sequences into templates: \textit{``If we can easily classify the blocks and make templates of the exploration order out of them, I think it would allow quick, adaptive ideation for many topics with just one click.''}

\begin{blockBig}
\textbf{Design Implication:} Systems should help users recognize and formalize their exploration strategies as path templates. These templates can serve as reusable scaffolds, allowing designers to quickly reapply their own strategies while adapting content to new contexts.
\end{blockBig}

\subsubsection{Project-level Reuse: Social Reuse as Collective Exploration}

At the project level, \sysname{} allowed participants to import either their own previous projects or those created by others. We observed two distinct purposes for this: (1) \textit{bootstrapping} and (2) \textit{broadening options}. 

For bootstrapping, participants reused the project at the beginning of exploration to avoid a blank start and quickly establish a foundation to build upon. In these cases, they tended to prefer smaller projects or their own prior project, and when reusing others' work, they selected projects whose intent was easy to understand so they could grasp the overall flow and extend it meaningfully. Because of this, participants valued \sysname{}’s structured representation for making the original designer’s reasoning legible. D1 reflected, \textit{``At first, I resisted using someone else's project because it wasn't my flow of thought. But once I tried, I could infer their thought process and work on top of that.''}

For broadening options, participants imported projects when they felt stuck and needed inspiration. In these cases, even when importing an entire project, participants treated it more like block- or path-level reuse --- borrowing intermediate intent units or recurring strategies rather than the entire process.
Interestingly, although \sysname{} provided adaptive reuse and recommendation features, participants overwhelmingly preferred exploring others' projects. D4 explained, \textit{``Compared to generated options which could be random, others' projects felt more valuable because those people must have actually found good things during their exploration.''}. This highlights how the combination of generative possibilities with human curation (embodied in others' projects) made them feel more trustworthy and valuable than purely algorithmic suggestions.

\begin{blockBig}
% bootstraping 목적으로 사람들이 사용할 때는 전반적인 과정이 잘 이해되도록 represent하는 게 중요하고, broadening 목적으로 사용할 때는 사람들의 키워드나 strategy 이런 요소들을 따로 따와서 browse 할 수 있도록
\textbf{Design Implication:} 
Systems that support sharing and reusing others’ exploration processes should treat the two purposes of reuse differently. For bootstrapping, systems should clearly represent the overall flow of intent evolution so users can build on it. For broadening options, systems should foreground intermediate steps, keywords, and strategies instead of only full processes or the final outputs, to allow selective browsing.
\end{blockBig}

% project-level reuse에는 (reuse 시기에 따른) 두 가지 목적이 있다
% (1) bootstrapping : "처음 시작할 때" blank state에서 시작하지 않고 뭔가 시작점을 구축해두려는?
% - 그래서 보통 작은 사이즈를 고름 + (내가 이전에 한 거 고르거나 or 다른 사람들 것 중에서도 의도가 명확히 이해되는 것 고름)
% - 그래서 ideablocks에서 전체 흐름이 보이는 그런 representation을 value함 -- 과정을 이해해야 그 위에서 일할 수 있으니까
% - D1 quote: "처음에는 다른 사람 걸 가져오면 내 생각의 흐름이 아니라서 거부감이 있었는데, 막상 써보니까, 그 사람의 여러 생각의 흐름이 유추가 되었고,(...그 위에서 더 탐색할 수 있었다)"
% (2) broadening options: "생각이 잘 안날 때" inspiration을 얻기 위한 옵션을 보기 위해서
% - 그래서 이 때는 프로젝트 전체를 import 해오더라도 block-level reuse나 path-level reuse랑 비슷하게 쓴다. (intermediary intent unit or strategy)
% - interestingly, 비슷한 inspiration을 system의 recommendation이나 adaptive한 옵션을 보면서도 얻을 수 있지만, 사람들은 압도적으로 다른 사용자의 과정을 보는 것을 선호했다 -- because of the trust on others' 고뇌? in their processes

\subsubsection{Reuse and Creativity: Balancing Divergence and Convergence}

Because intent reuse involves drawing on past actions, some participants expressed concern that it might steer exploration toward over-convergence.

At the same time, others highlighted how reuse could also spark more divergence and personal expression.
Participants noted that the adaptive reuse options helped them sustain divergence even when reusing. D5 explained, \textit{``Reusing elements didn't feel like convergent as when I copied the same one, I still saw other variations, and those adaptive options kept it fresh.''}
For literal reuse, several participants reframed it not as convergence but as a way to build one's own identity. D4 noted, \textit{``Designers develop their own style. Saving and reusing that style is not a loss of originality --- it's part of constructing identity.''}

Participants also suggested system-level nudges to counter over-reliance on recent or popular items. D6 proposed, \textit{``I could imagine the system blurs the recently reused blocks or paths so that older ones resurfaced, and it would encourage me to reuse in more diverse ways.''} Extending this to social reuse, de-emphasizing overly popular projects or ideas could help avoid collective homogeneity and encourage discovery of less obvious alternatives.

\begin{blockBig}
\textbf{Design Implication:} Systems should offer adaptive variants alongside literal reuse to sustain divergence. Also, the system could introduce nudges such as recency fading to prevent over-reliance on the same items.
\end{blockBig}

%% file: Sections/08-Discussion.tex
\section{Discussion}

\subsection{Divergent Intent as a Parameterizable Construct}\label{sec:generalizability}
In \sysname{}, we conceptualized \textbf{\textit{divergent intent}} not merely as a vague desire for alternatives, but as the explicit articulation of a parametric search space. While current generative AI models primarily optimize for convergent intent --- refining output toward a specific, fixed target --- early-stage design requires the flexibility to define and manipulate the boundaries of exploration. \sysname{} operationalizes this by decomposing intent into (property, direction, range), and providing a pipeline and interface to parametrically tune it. By shifting the interaction mechanism from instruction (telling the model ``what'' to generate) to sculpting (defining ``where'' and ``how'' to search), we transformed the generative model into a controllable engine for divergent thinking, allowing designers to structure the ill-defined design space.

While \sysname{} targets the visually open-ended nature of graphic design, its core mechanism of parameterizing divergent exploration into \textit{Property}, \textit{Direction}, and \textit{Range} can be effectively adapted to other domains by redefining the specific parameters available for manipulation.

Across design tasks, the specific instantiation of these elements may vary. For example, while graphic design commonly foregrounds properties such as `image style' or `character entity', other domains may prioritize different aspects, such as `layout' in UI/UX design or `materials' in 3D product design.
Similarly, the meaning and operationalization of \textit{range} may vary across tasks. In \sysname{}, we instantiate range using \textit{typicality} to capture how far exploration deviates from a familiar center. In other contexts, designers may instead wish to control parameters such as abstraction (literal to symbolic), visual fidelity (sketchy to polished), or constraint strength (loose to strict) to better align exploration with their own concerns. By externalizing divergent intent into explicit, task-appropriate parameters, future systems can adapt this approach to support exploratory needs across a wider range of creative practices.

\subsection{Implications for Design: Scaffolding and Sustaining Divergence}
\sysname{} demonstrates how modularizing and reusing divergent intent helps designers to explore better. In this section, we further discuss the future opportunities for intent-aware creativity support systems.

Regarding intent expression, our findings suggest a trade-off between explicit scaffolding and tacit fluency. \sysname{} lowers barriers for early-stage designers by externalizing the mental model of exploration (i.e., Property, Direction, Range). For novices who lack intuitive strategies to structure their search, the interface acts as a pedagogical scaffold to help them navigate the design space.
However, this explicit parameterization may introduce friction for experts who rely on tacit intuition~\cite{polanyi2009tacit, son2024demystifying}. To support expert users, future systems could be improved to infer users' tacit intent rather than requiring explicit input. For example, the system can capture tacit signals such as dwelling time on specific variations or spontaneous verbal reactions to automatically tune exploration parameters. This approach would allow the system to adaptively shape the search space without interrupting the designer's intuitive workflow.

Regarding intent reuse, participants from Study 2 noted the risk that reuse might lead to homogenization or over-convergence. While \sysname{} addresses this by offering adaptive reuse options, which mutate the original intent to fit new contexts, we suggest that future systems must proactively intervene to prevent the formation of this echo chamber. To mitigate reliance on familiar patterns, systems could introduce algorithmic nudges, for example, by temporarily de-emphasizing recently or frequently used blocks, allowing older or less popular strategies to resurface. By artificially diversifying the visibility of reusable intents, the system can disrupt potential fixation loops and encourage designers to adopt a diverse range of strategies.

\subsection{The Diversity-Creativity Gap}\label{sec:convergence}
\sysname{} was designed to isolate and strengthen the divergent phase of the creative process, enabling clearer examination of how such support impacts early-stage expansion of the design space. Within this scope, \sysname{} increased the breadth, structure, and continuity of exploration. However, these gains did not translate into significantly higher overall creativity ratings, suggesting that divergent exploration alone does not fully account for perceived creativity.

One explanation concerns how creative value is measured and experienced. Although \sysname{} accelerated the generation of options, this efficiency does not inherently lead to higher creativity. Faster exploration can reduce the sense of authorship if transitions toward deliberate selection and refinement are not well supported, suggesting that objective diversity metrics capture only one dimension of creative work.

Another explanation relates to the role of convergent processes.
While \sysname{} includes lightweight mechanisms that allow certain aspects of an exploration to be locked, these features do not constitute comprehensive support for convergent refinement. P8 stated, \textit{``When I don't have a clear idea in mind, I like the broadness of system suggestions. But once I get a specific vision, it's difficult to control each block precisely and get the exact image that I want.''}
This tension between divergence and convergence is further reflected in the baseline condition: baseline participants may have reached a satisfactory result more quickly and stopped exploring, such that their lower ideational fluency reflects efficient convergence rather than a lack of divergent capability.
Together, this suggests that creativity ratings may be shaped less by the volume of exploration than by how well the system supports the full arc from divergence to convergence.

From this perspective, future work can build on \sysname{}'s divergent intent supports by enabling more fluid transitions between divergent and convergent intent. One promising direction is to allow exploratory outcomes (such as stabilized block or paths) to function as intermediate representations that inform subsequent convergent activities, either within the same system or through other tools. Supporting such connections iteratively would help preserve the value of divergent exploration as designers move toward more focused refinement.

%% file: Sections/09-Conclusion.tex
\section{Conclusion}

This paper presents \sysname{} to support the divergent phase of design exploration by operationalizing divergent intent into reusable units comprised of property, direction, and range. By modularizing these intents, \sysname{} allows designers not only to shape the boundary of exploration but also to reuse the exploration logic across different contexts.
Our evaluation showed that this structured approach significantly increases the diversity and connectivity of exploration. A longitudinal study revealed how designers appropriated reuse features in distinct ways --- block reuse as palettes, path reuse as templates, and project reuse for bootstrapping or inspiration. These observations lead to design implications for future intent-reusing creativity support systems.

%% file: Sections/99-Appendix.tex
\appendix
\onecolumn
\section{Formative Study}
\subsection{Details of Formative Study Participants} \label{appendix:formativeparticipant}

Details of the formative study participants are shown in Table~\ref{tab:formativeparticipant}. Participants responded to two 7-point Likert scale questions asking about their familiarity with generative AI ("How much do you understand how image generation AI models work?") in the enrollment survey.

\begin{table*}[h]
\begin{tabular}{cccccc}
\toprule
\textbf{PID} & \textbf{Age} & \textbf{Sex} & \textbf{Design Related Background}         & \textbf{\begin{tabular}[c]{@{}c@{}}Frequency of Using\\ Generative AI Model\end{tabular}} & \textbf{\begin{tabular}[c]{@{}c@{}}Generative AI\\ familiarity (7-scale)\end{tabular}} \\ \hline
P1           & 23           & F            & Undergrad student in design & Less than 1 time a month                                                                  & 3                                                                                      \\
P2           & 24           & F            & Master's student in design                 & Less than 1 time a month                                                                  & 3                                                                                      \\
P3           & 26           & F            & Master's student in design                 & Less than 1 time a month                                                                  & 3                                                                                      \\
P4           & 23           & F            & Worked as a freelancer designer            & Never                                                                                     & 2                                                                                      \\
P5           & 23           & F            & Undergrad student in design & Less than 1 time a month                                                                  & 3                                                                                      \\
P6           & 23           & F            & Bachelor's graduate in design              & 2-3 times a month                                                                         & 4                                                                                      \\
P7           & 32           & M            & Bachelor's graduate in design              & 1 time a week                                                                             & 4                                                                                      \\
\bottomrule
\end{tabular}
\caption{Participants’ demographic information, design-related background, frequency of generative AI model usage, and self-reported familiarity on a 7-point Likert scale.}
\Description{This table presents the demographic details, design-related backgrounds, and generative AI usage patterns of the seven participants (P1–P7) in the formative study. Participants ranged in age from 23 to 32 years, with 6 females and 1 male. All had some form of design background, including Undergrad and graduate students in design, bachelor's graduates, and one freelance designer. Their experience with generative AI varied: most used such models less than once a month or not at all, and self-reported familiarity levels ranged from 2 to 4 on a 7-point Likert scale.}
\label{tab:formativeparticipant}
\end{table*}

\subsection{Thematic Analysis Codebook}\label{appendix:codebook}
This codebook outlines the three components of divergent intent that emerged from our inductive thematic analysis, with example quotes from the think-aloud sessions where participants articulated their divergent intent, the interaction experiences that contributed to each component, and friction points identified in interviews.

\begin{table*}[h]
\begin{tabular}{@{}%
  L{0.08\textwidth}  % Theme
  L{0.17\textwidth}  % Definition
  L{0.22\textwidth}  % Representative Excerpts
  L{0.20\textwidth}  % Contributing Interaction Experience
  L{0.25\textwidth}  % Associated Pain Points
@{}}
\toprule
\textbf{Theme} &
\textbf{Definition} &
\textbf{Representative Quotes} &
\textbf{Interactions} &
\textbf{Associated Pain Points} \\ \midrule
\textbf{Property} &
The specific design attribute (e.g., Subject, Style, Composition) to focus on. &
\textit{"I like the character, just change the background style." (P7) \newline
"Keep the composition fixed, but try different rendering styles." (P4)} &
Specifying which part of the prompt to modify, or through inpainting to isolate a specific design attribute &
- Users cannot isolate specific properties e.g., changing the style prompt unintentionally alters the subject or composition \\ \midrule
\textbf{Direction} &
The qualitative vector or ``vibe'' the user wishes to steer towards. &
\textit{"I want to see something like this reference, but with a more `sketchy' feel." (P5) \newline
"I'm looking for a `cyberpunk' atmosphere, but less aggressive." (P2)} &
Primarily through text prompting; occasionally through reference-based image-to-image generation &
- Hard to verbalize the visual direction \newline - When users want to update the direction after looking at the results, they struggle how to change them \\ \midrule
\textbf{Range} &
The magnitude of variance; the scope of exploration from the anchor. &
\textit{"I want to see totally different layout options, not just small tweaks." (P1) \newline
"Keep it mostly the same, just surprise me a little bit with the details." (P3)} &
Image variation features and prompt modifiers (e.g., ``various styles''), though these were often unreliable &
- Users had hard time express how much the model deviates in a structured way \newline - Reliance on trial-and-error or prompt hacking to generate the right level of diversity \\
\bottomrule
\end{tabular}
\caption{Thematic Mapping of Divergent Intent Components, Contributing Interaction Experiences, and Associated Pain Points.}
\Description{This table shows the thematic analysis codebook which maps the three components of divergent intent—Property, Direction, and Range—to their definitions, representative quotes, contributing interaction experiences, and associated pain points. Property refers to specific attributes (e.g., Style), emerging from prompt editing and inpainting, where users struggle to isolate changes without altering other elements. Direction covers the qualitative ``vibe,'' primarily arising from text prompting and reference-based generation, noting difficulties in verbalizing or updating visual goals. Range defines the magnitude of variance, highlighted through image variation features and prompt modifiers, with a lack of structured control forcing users to rely on trial-and-error.}
\end{table*}

\section{Pipeline Design}

\subsection{Image Generation Pipeline}\label{appendix:tech}
To generate images that incorporate both previous exploration history and the user's new divergent intent, \sysname{} composes prompts using a two-step LLM pipeline. First, it prompts the LLM to extract detailed descriptions of previously explored properties from the most recent image in the connected block sequence. Then, it generates descriptions about the user's new exploration intent: for text-based properties, it uses the provided suggestion directly; for image-based properties, it prompts GPT-4o to produce a description of the new visual concept. GPT-4o then synthesizes these elements into a text-to-image prompt, which is passed to DALL-E 3 to generate the final image.

With the release of GPT-image-1, we simplified this pipeline using its image-editing model. The model takes the most recent image, its explored properties, and the user's new intent as input, blending the new property into the existing image while preserving previously established visual characteristics. The full prompts used in this pipeline are provided 
in the supplementary materials.

\section{Evaluation}

\subsection{Baseline System}

\begin{figure*}[h]
    \centering
    \includegraphics[width=0.83\linewidth]{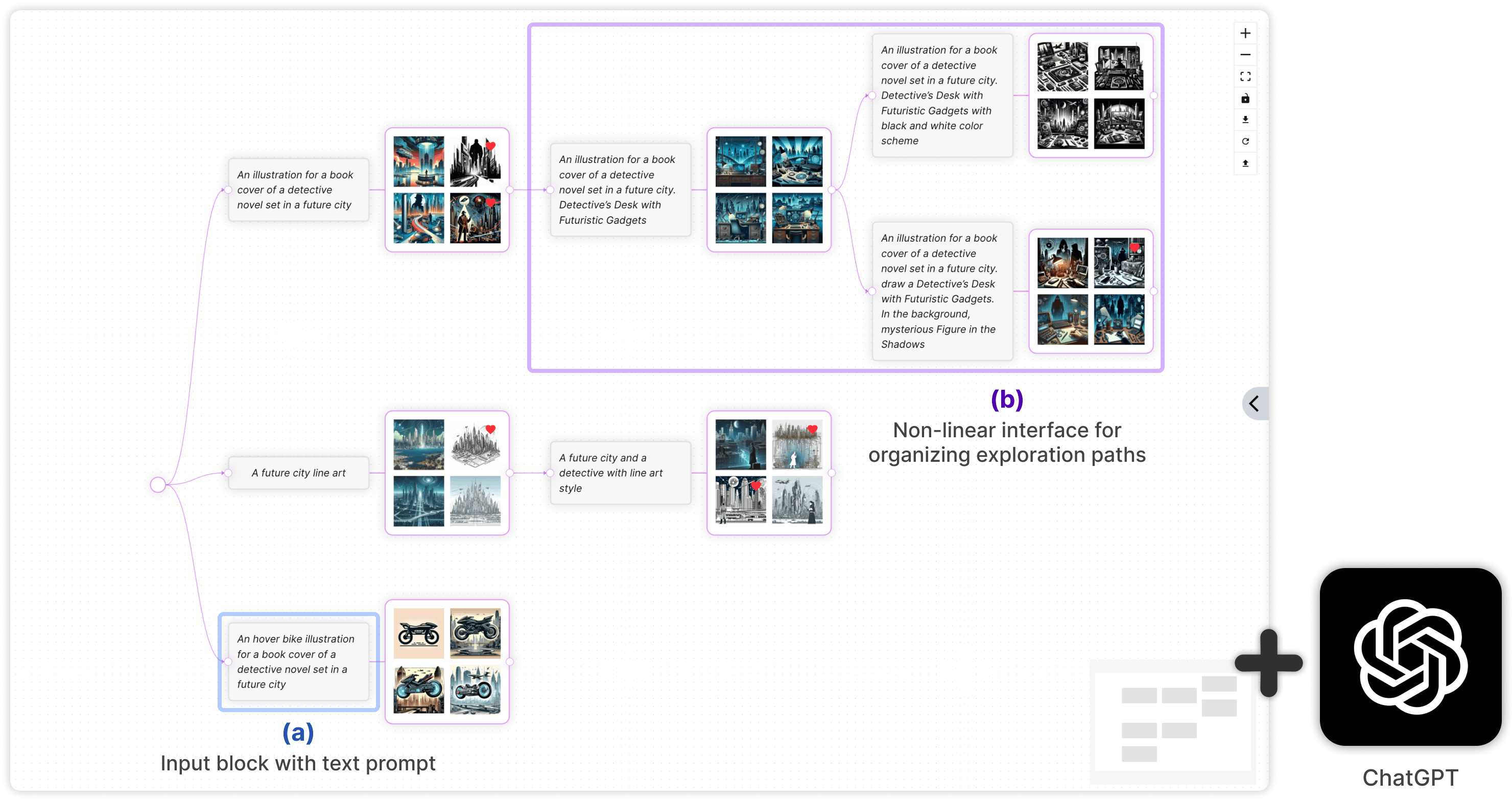}
    \caption{Screenshot of the baseline system. (a) Users initiate exploration by entering free-form text prompts. (b) The canvas-based interface supports non-linear branching of ideas, but without features for structured intent expression or reuse, such as direction/range inputs or path reuse. Users can use ChatGPT externally for assistance during exploration.}
    \Description{
    A screenshot of the baseline system's interface. On the left, labeled (a), users input a free-form text prompt to generate image outputs. These prompts are organized into a branching structure on a non-linear canvas, shown on the right. (b) Each prompt node leads to a set of four generated images. Although branching is supported visually, it does not preserve structured intent or offer chaining mechanisms like in \sysname{}. No structured UI elements like sliders or block types are present. The ChatGPT logo in the bottom-right indicates that users may consult ChatGPT separately.
    }
    \label{fig:screenshot-baseline}
\end{figure*}

\subsection{Details of Study Participants} \label{appendix:studyparticipant}

\begin{table*}[h]
\begin{tabular}{ccccccc}
\toprule
\textbf{PID} & \textbf{Age} & \textbf{Sex} & \textbf{Design Related Background}    & \textbf{\begin{tabular}[c]{@{}c@{}}Frequency of Using\\ Generative AI Model\end{tabular}} & \textbf{\begin{tabular}[c]{@{}c@{}}Familiarity of\\using GenAI\end{tabular}} & \textbf{\begin{tabular}[c]{@{}c@{}}Understanding\\of GenAI\end{tabular}} \\ \hline
P1           & 24           & F            & Undergrad student in design       & 1–2 times a month                                                                         & 5                                                                                             & 6                                                                                         \\
P2           & 22           & M            & Undergrad student in design       & Once a week                                                                               & 5                                                                                             & 5                                                                                         \\
P3           & 26           & M            & Ph.D. student in design               & Once in 2-3 months                                                                        & 3                                                                                             & 3                                                                                         \\
P4           & 28           & M            & Ph.D. student in design               & 1–2 times a month                                                                         & 4                                                                                             & 5                                                                                         \\
P5           & 26           & M            & Ph.D. student in design               & 1–2 times a month                                                                         & 6                                                                                             & 4                                                                                         \\
P6           & 26           & M            & Master's student in design            & Once a week                                                                               & 4                                                                                             & 6                                                                                         \\
P7           & 23           & F            & Master's student in design            & 2-5 times a week                                                                          & 4                                                                                             & 3                                                                                         \\
P8           & 25           & F            & Bachelor's graduate in design         & 1–2 times a month                                                                         & 4                                                                                             & 5                                                                                         \\
P9           & 25           & M            & Undergrad student in design$^*$       & Once a week                                                                               & 5                                                                                             & 5                                                                                         \\
P10          & 24           & F            & Worked as a freelancer designer       & 1–2 times a month                                                                         & 5                                                                                             & 2                                                                                         \\
P11          & 27           & M            & Master's graduate in design           & 2-5 times a week                                                                          & 6                                                                                             & 4                                                                                         \\
P12          & 24           & F            & Master's student in design            & Once in 2-3 months                                                                        & 4                                                                                             & 4                                                                                         \\ \hline
D1           & 27           & F            & Ph.D. student in design               & 2-5 times a week                                                                          & 6                                                                                             & 6                                                                                         \\
D2           & 25           & F            & Undergrad student in design       & 2-5 times a week                                                                          & 5                                                                                             & 3                                                                                         \\
D3          & 25           & M            & UX/UI designer (2 years)$^*$       & 2-5 times a week                                                                          & 5                                                                                             & 4                                                                                         \\
D4           & 27           & M            & Game designer (3 years)               & Once a week                                                                               & 6                                                                                             & 5                                                                                         \\
D5           & 25           & F            & Freelancer graphic designer (2 years) & 2-5 times a week                                                                          & 5                                                                                             & 5                                                                                         \\
D6           & 19           & M            & UX/UI designer (6 months)             & Almost every day                                                                          & 6                                                                                             & 5                                                                                         \\
\bottomrule
\end{tabular}
\\[4pt]
{\footnotesize $^*$ P9 and D3 are the same participant, who transitioned from student to professional between the two studies.}
\caption{Participant details for the comparison study (P1–P12) and longitudinal study (D1–D6), including age, sex, design background, GenAI usage frequency, and self-reported familiarity and understanding of generative AI (7-point Likert scale).}
\Description{This table presents the demographic details, design-related backgrounds, and generative AI experience of the 12 participants (P1-P12) in the comparison study and 6 participants (D1-D6) in the longitudinal study. Participants ranged in age from 19 to 28 years, including 8 females and 10 males. All participants had a design background, with roles such as Undergrad and graduate students in design, Ph.D. students, bachelor's graduates, and professional designers with 6 months to 3 years of experience. Their frequency of using generative AI varied from once in 2–3 months to almost every day. Participants also rated their familiarity with and understanding of generative AI models on a 7-point Likert scale. Familiarity scores ranged from 3 to 6, while understanding scores ranged from 2 to 6.}
\label{tab:participant}
\end{table*}

\subsection{Comparative Study Interview questions}\label{appendix:interviewquestions}
These are the questions used for the semi-structured interview after the two design exploration sessions with baseline and \sysname{}.

\subsubsection*{\sysname{} vs. Baseline}

\begin{itemize}
    \item What was the biggest difference when exploring design ideas using the two tools?
    \item Was there a difference in the way you explored design ideas when using the two tools?
\end{itemize}

\subsubsection*{Intent Expression}

\begin{itemize}\item Was there a difference in how you conveyed your exploration intent when using the two tools? (e.g., text prompt vs. structured input)
    \item Which tool was better for fostering more creative/efficient exploration?
\end{itemize}

\subsubsection*{Revisiting \& Intent Evolution}

\begin{itemize}
    \item How did your exploration intent change when using the two tools?
    \item During the exploration process, did you continually refer to your previous results?
    \begin{itemize}
        \item If so, how did your approach to referencing previous results differ when using each tool?
    \end{itemize}
\end{itemize}

\subsubsection*{\sysname{} / Baseline Specific}

\begin{itemize}
    \item What was the most useful feature of \sysname{}? In what cases was it particularly helpful?
    \item What was the least used feature of \sysname{}? Why did you use it less?
    \item In what cases and how did you use ChatGPT (in baseline condition)?
\end{itemize}

\subsubsection*{Comparison with Usual Design Exploration Methods}

\begin{itemize}
    \item Was the design exploration process using the provided tools similar to your usual method of exploring design ideas? If there were differences, please explain.
    \item The most significant difference might be the non-linear interface. How was your experience with that?
\end{itemize}

\subsection{Longitudinal Study Interview questions}\label{appendix:interviewquestions2}
These are the questions used for the semi-structured interview after the two design exploration sessions with baseline and \sysname{}.

\subsubsection*{Overall Experience}

\begin{itemize}
    \item How was your overall experience after using the system for three days?
    \item Compared to the way you usually explore design (e.g., Google Image Search, Pinterest, etc.), what felt most different?
    \item Compared to the way you usually use image generation models, what felt most different?
\end{itemize}

\subsubsection*{Experience with Intent Reusing Feature}

\begin{itemize}
    \item Our system provided three ways to reuse previous explorations (Block, Path, Project). Which of the three did you use the most?
    \item In what situations did you use each unit of the reusing?
    \item When reusing a Project, did you ever start by referencing someone else’s project? If so, what aspects did you usually reference? Was it easy to understand their exploration process?
    \item Each time you reused an exploration, our system provided two results (literal, adaptive). Which one did you prefer?
    \item Were the two approaches useful in different situations? If so, how did they differ?
\end{itemize}

\subsubsection*{Differences Depending on Context}

\begin{itemize}
    \item Did your exploration approach change between the early and later stages of completing a task? If yes, how?
    \item From Day 1 to Day 3, did your way of using the system evolve over time? If yes, how?
    \item Did the familiarity of the task affect how you explored? If yes, how?
    \item Were there any other factors that influenced the way you used the system?
\end{itemize}

\subsubsection*{Future Improvement}

\begin{itemize}
    \item Did you feel that the reuse feature supported creative exploration, or did it make the results converge instead?
    \item In what ways do you wish the system could better support your personal working style?
\end{itemize}

\section{Results}

\subsection{Example Linkographs from User Study Sessions}
\begin{figure*}[h]
    \centering
    \includegraphics[width=1\linewidth]{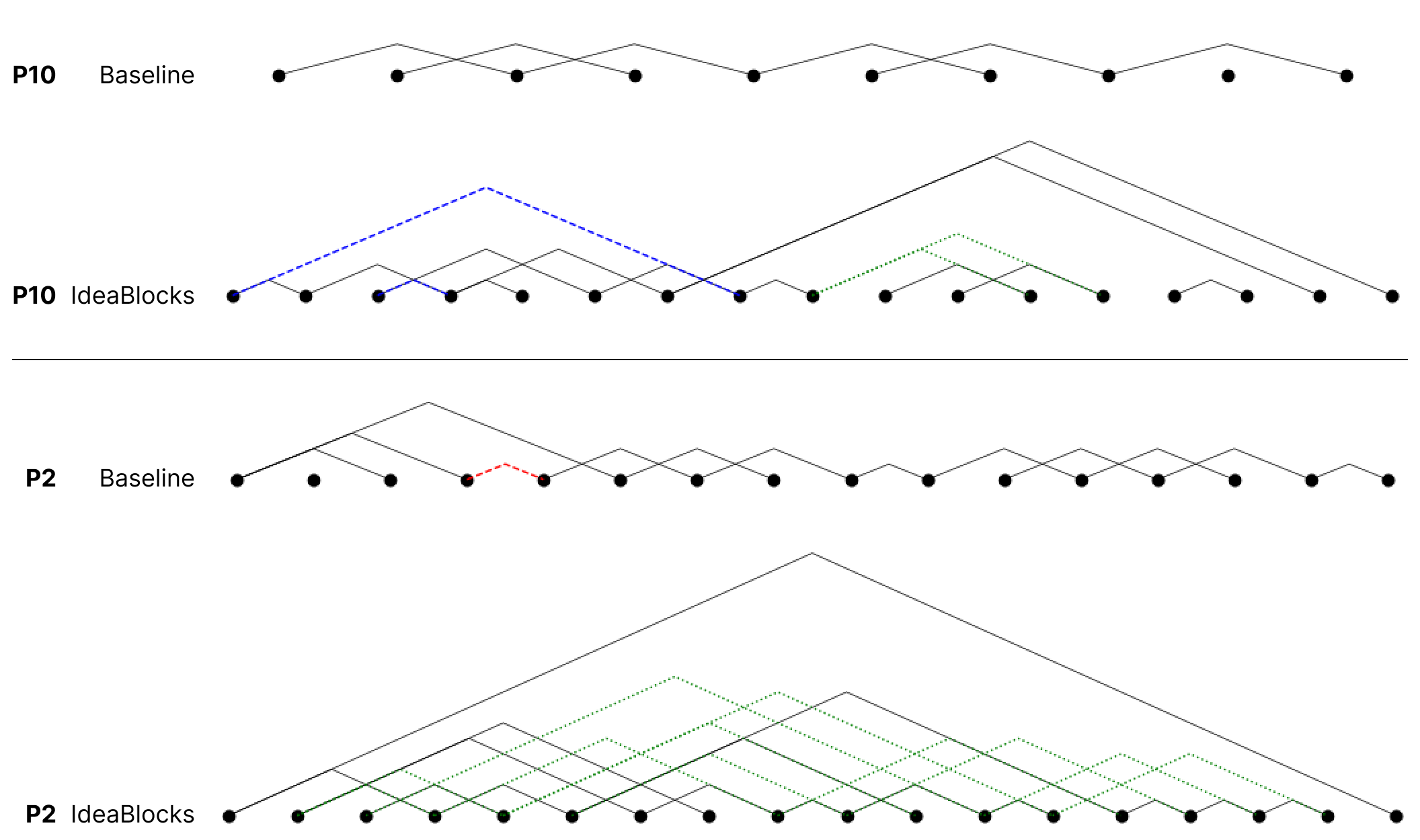}
    \caption{Example linkographs from two participants (P10 and P2) in both baseline and \sysname{} conditions. Each node represents an input block created during ideation. Solid black lines indicate parent-child connections, blue dashed lines indicate history-based reuse, green dotted lines indicate copy-based reuse, and red dotted lines indicate manually identified references based on think-aloud data. Compared to the baseline, participants using \sysname{} exhibited more frequent reuse and more interconnected idea structures.}
    \Description{
    The figure shows linkographs for two participants (P10 and P2), with separate diagrams for the baseline and \sysname{} conditions. Each diagram contains a horizontal sequence of black dots, representing individual input blocks. Black solid lines connect adjacent dots to show parent-child relationships. In the \sysname{} diagrams, additional dotted lines appear: blue dashed lines indicate links created through reusing past history, green dotted lines represent blocks reused by copying, and a red dotted line (in P2's baseline) marks a manually identified reference from think-aloud data. The \sysname{} linkographs display more branching and crisscrossing lines, showing more interconnected and reuse-rich exploration paths compared to the more linear baseline structures.
    }
    \label{fig:linkograph-example}
\end{figure*}

To further illustrate differences in exploration structure, Figure~\ref{fig:linkograph-example} shows qualitative examples of linkographs from participants' sessions in both conditions.

\subsection{CSI \& NASA-TLX Survey Results} \label{appendix:fullsurvey}
Table~\ref{tab:fullsurvey} shows the survey results for NASA-TLX and Creativity Support Index (CSI) between \sysname{} and baseline.

\begin{table*}[h]
\begin{tabular}{ccrrrrr}
\toprule
                                                                                          &                      & \multicolumn{2}{c}{\sysname{}}                 & \multicolumn{2}{c}{baseline}                    & \multicolumn{1}{c}{\multirow{2}{*}{$p$}} \\ \cline{3-6}
                                                                                          &                      & \multicolumn{1}{c}{$M$} & \multicolumn{1}{c}{$SD$} & \multicolumn{1}{c}{$M$} & \multicolumn{1}{c}{$SD$} & \multicolumn{1}{c}{}                   \\ \hline
\multirow{6}{*}{\begin{tabular}[c]{@{}c@{}}Creativity Support Index\\(CSI)\end{tabular}} & Enjoyment            & 6.17                  & 0.85                    & 5.38                  & 1.49                    & 0.1397                                 \\
                                                                                          & Exploration          & 6.00                  & 0.87                    & 5.33                  & 0.90                    & 0.0901                                 \\
                                                                                          & Expressiveness       & 5.79                  & 0.75                    & 5.00                  & 1.26                    & 0.0867                                 \\
                                                                                          & Immersion            & 4.63                  & 1.87                    & 4.42                  & 1.92                    & 0.7993                                 \\
                                                                                          & Results Worth Effort & 5.21                  & 1.31                    & 5.00                  & 1.10                    & 0.6906                                 \\
                                                                                          & Collaboration        & 5.63                  & 0.77                    & 5.17                  & 1.57                    & 0.3943                                 \\ \hline
\multirow{6}{*}{NASA-TLX}                                                                 & Mental demand        & 2.75                  & 1.69                    & 2.92                  & 1.66                    & 0.4314                                 \\
                                                                                          & Physical demand      & 1.75                  & 1.42                    & 1.92                  & 1.44                    & 0.1573                                 \\
                                                                                          & Temporal demand      & 2.08                  & 1.19                    & 2.25                  & 1.42                    & 0.7505                                 \\
                                                                                          & Effort               & 2.25                  & 1.01                    & 2.75                  & 1.42                    & 0.3008                                 \\
                                                                                          & Performance          & 5.08                  & 1.19                    & 4.67                  & 1.43                    & 0.4657                                 \\
                                                                                          & Frustration          & 1.92                  & 1.04                    & 2.42                  & 1.38                    & 0.1573                                 \\ 
\bottomrule
\end{tabular}
\caption{Survey results from NASA-TLX and Creativity Support Index (CSI) on a 7-point Likert scale. Each value shows the mean ($M$) and standard deviation ($SD$).}
\Description{This table shows the survey results comparing the proposed system and a baseline system using the Creativity Support Index (CSI) and NASA-TLX, both measured on a 7-point Likert scale. For CSI, participants rated higher scores for the proposed system across all subscales, although none of the differences were statistically significant. For NASA-TLX, both systems scored low on physical and mental demand. Participants reported slightly lower effort and frustration and slightly better performance when using the proposed system, although none of the differences were statistically significant.}
\label{tab:fullsurvey}
\end{table*}

% \clearpage
% \input{Sections/99-Prompts}